\documentclass[a4paper,12pt, epsfig]{article}
\usepackage{epsfig}
\usepackage{amssymb,latexsym}
\usepackage{amsfonts}
\usepackage{amsmath}
\usepackage[colorlinks, linkcolor=blue, citecolor=blue, urlcolor=blue]{hyperref}
\usepackage{epstopdf}

\newskip\humongous \humongous=0pt plus 1000pt minus 1000pt

\newif\ifdtup

\catcode`\@=11

\@addtoreset{equation}{section}

\def\@normalsize{\@setsize\normalsize{15pt}\xiipt\@xiipt
\abovedisplayskip 14pt plus3pt minus3pt%
\belowdisplayskip \abovedisplayskip
\abovedisplayshortskip \z@ plus3pt%
\belowdisplayshortskip 7pt plus3.5pt minus0pt}

\def\small{\@setsize\small{13.6pt}\xipt\@xipt
\abovedisplayskip 13pt plus3pt minus3pt%
\belowdisplayskip \abovedisplayskip
\abovedisplayshortskip \z@ plus3pt%
\belowdisplayshortskip 7pt plus3.5pt minus0pt
\def\@listi{\parsep 4.5pt plus 2pt minus 1pt
      \itemsep \parsep
      \topsep 9pt plus 3pt minus 3pt}}

\relax

\catcode`@=12

\catcode`\@=11

\def\section{\@startsection{section}{1}{\z@}{3.5ex plus 1ex minus
    .2ex}{2.3ex plus .2ex}{\large\bf}}

\def\thesection{\arabic{section}}
\def\thesubsection{\arabic{section}.\arabic{subsection}}

\def\appendix{\setcounter{section}{0}
  \def\thesection{Appendix \Alph{section}}
  \def\thesubsection{\Alph{section}.\arabic{subsection}}
  \def\theequation{\Alph{section}.\arabic{equation}}}

\def\SymBoxes#1#2#3#4{\newdimen\un@t \un@t#3%
\raisebox{#1}{\rule{#2\un@t}{#4}\hskip-#2\un@t
\@tempdimb\un@t \advance\@tempdimb by-#4\@tempcntb#2\relax%
\@whilenum{\@tempcntb>0}\do{
\rule{#4}{\un@t}\hskip\@tempdimb \advance\@tempcntb by\m@ne}%
\hskip-#2\un@t \rule[\un@t]{#2\un@t}{#4}%
\rule[\un@t]{#4}{#4}\hskip-#4
\rule{#4}{\un@t}}\hskip-#4}                

\begin{document}


\newcommand{\dd}{\textrm{d}}

\newcommand{\beq}{\begin{equation}}
\newcommand{\eeq}{\end{equation}}
\newcommand{\bea}{\begin{eqnarray}}
\newcommand{\eea}{\end{eqnarray}}
\newcommand{\beas}{\begin{eqnarray*}}
\newcommand{\eeas}{\end{eqnarray*}}
\newcommand{\defi}{\stackrel{\rm def}{=}}
\newcommand{\non}{\nonumber}
\newcommand{\bquo}{\begin{quote}}
\newcommand{\enqu}{\end{quote}}
\newcommand{\tc}[1]{\textcolor{blue}{#1}}
\renewcommand{\(}{\begin{equation}}
\renewcommand{\)}{\end{equation}}
\def\de{\partial}
\def\Om{\ensuremath{\Omega}}
\def\Tr{ \hbox{\rm Tr}}
\def\rc{ \hbox{$r_{\rm c}$}}
\def\H{ \hbox{\rm H}}
\def\HE{ \hbox{$\rm H^{even}$}}
\def\HO{ \hbox{$\rm H^{odd}$}}
\def\HEO{ \hbox{$\rm H^{even/odd}$}}
\def\HOE{ \hbox{$\rm H^{odd/even}$}}
\def\HHO{ \hbox{$\rm H_H^{odd}$}}
\def\HHEO{ \hbox{$\rm H_H^{even/odd}$}}
\def\HHOE{ \hbox{$\rm H_H^{odd/even}$}}
\def\K{ \hbox{\rm K}}
\def\Im{ \hbox{\rm Im}}
\def\Ker{ \hbox{\rm Ker}}
\def\const{\hbox {\rm const.}}
\def\o{\over}
\def\im{\hbox{\rm Im}}
\def\re{\hbox{\rm Re}}
\def\bra{\langle}\def\ket{\rangle}
\def\Arg{\hbox {\rm Arg}}
\def\exo{\hbox {\rm exp}}
\def\diag{\hbox{\rm diag}}
\def\longvert{{\rule[-2mm]{0.1mm}{7mm}}\,}
\def\a{\alpha}
\def\b{\beta}
\def\e{\epsilon}
\def\l{\lambda}
\def\ol{{\overline{\lambda}}}
\def\ochi{{\overline{\chi}}}
\def\th{\theta}
\def\s{\sigma}
\def\oth{\overline{\theta}}
\def\ad{{\dot{\alpha}}}
\def\bd{{\dot{\beta}}}
\def\oD{\overline{D}}
\def\opsi{\overline{\psi}}
\def\dag{{}^{\dagger}}
\def\tq{{\widetilde q}}
\def\L{{\mathcal{L}}}
\def\p{{}^{\prime}}
\def\W{W}
\def\N{{\cal N}}
\def\hsp{,\hspace{.7cm}}
\def\hspp{,\hspace{.5cm}}
\def\bo{\ensuremath{\hat{b}_1}}
\def\bfo{\ensuremath{\hat{b}_4}}
\def\co{\ensuremath{\hat{c}_1}}
\def\cfo{\ensuremath{\hat{c}_4}}
\def\th#1#2{\ensuremath{\theta_{#1#2}}}
\def\c#1#2{\hbox{\rm cos}(\th#1#2)}
\def\s#1#2{\hbox{\rm sin}(\th#1#2)}
\def\cp#1#2#3{\hbox{\rm cos}^#1(\th#2#3)}
\def\sp#1#2#3{\hbox{\rm sin}^#1(\th#2#3)}
\def\ctp#1#2#3{\hbox{\rm cot}^#1(\th#2#3)}
\def\cpp#1#2#3#4{\hbox{\rm cos}^#1(#2\th#3#4)}
\def\spp#1#2#3#4{\hbox{\rm sin}^#1(#2\th#3#4)}
\def\t#1#2{\hbox{\rm tan}(\th#1#2)}
\def\tp#1#2#3{\hbox{\rm tan}^#1(\th#2#3)}
\def\m#1#2{\ensuremath{\Delta M_{#1#2}^2}}
\def\mn#1#2{\ensuremath{|\Delta M_{#1#2}^2}|}
\def\u#1#2{\ensuremath{{}^{2#1#2}\mathrm{U}}}
\def\pu#1#2{\ensuremath{{}^{2#1#2}\mathrm{Pu}}}
\def\meff{\ensuremath{\Delta M^2_{\rm{eff}}}}
\def\an{\ensuremath{\alpha_n}}
\newcommand{\C}{\ensuremath{\mathbb C}}
\newcommand{\Z}{\ensuremath{\mathbb Z}}
\newcommand{\R}{\ensuremath{\mathbb R}}
\newcommand{\rp}{\ensuremath{\mathbb {RP}}}
\newcommand{\vac}{\ensuremath{|0\rangle}}
\newcommand{\vact}{\ensuremath{|00\rangle}                    }
\newcommand{\oc}{\ensuremath{\overline{c}}}
\renewcommand{\cos}{\textrm{cos}}
\renewcommand{\sin}{\textrm{sin}}
\renewcommand{\cot}{\textrm{cot}}

\newcommand{\Vol}{\textrm{Vol}}

\newcommand{\half}{\frac{1}{2}}

\def\changed#1{{\bf #1}}

\begin{titlepage}

\def\thefootnote{\fnsymbol{footnote}}

\begin{center}
{\large {\bf
The High Energy Neutrino Nuisance\\ at a Medium Baseline Reactor Experiment
  } }

\bigskip

\bigskip

{\large \noindent Emilio
Ciuffoli$^{1}$\footnote{ciuffoli@ihep.ac.cn}, Jarah
Evslin$^{1}$\footnote{\texttt{jarah@ihep.ac.cn}} and Xinmin Zhang$^{2,
1}$\footnote{\texttt{xmzhang@ihep.ac.cn}} }
\end{center}

\renewcommand{\thefootnote}{\arabic{footnote}}

\vskip.7cm

\begin{center}
\vspace{0em} {\em  { 1) TPCSF, IHEP, Chinese Acad. of Sciences\\
2) Theoretical physics division, IHEP, Chinese Acad. of Sciences\\
YuQuan Lu 19(B), Beijing 100049, China}}

\vskip .4cm

\vskip .4cm

\end{center}

\vspace{1.3cm}

\noindent
\begin{center} {\bf Abstract} \end{center}

\noindent
10 years from now medium baseline reactor experiments will attempt to determine the neutrino mass hierarchy from the differences $(RL+PV)$ between the extrema of the Fourier transformed neutrino spectra.   Recently Qian et al. have claimed that this goal may be impeded by the strong dependence of the difference parameter $RL+PV$ on the reactor neutrino flux and on slight variations of $\mn32$.  We demonstrate that this effect results from a spurious dependence of the difference parameter on the very high energy (8+ MeV) tail of the reactor neutrino spectrum.  This dependence is spurious because the high energy tail depends upon decays of exotic isotopes and is insensitive to the mass hierarchy.  An energy-dependent weight in the Fourier transform not only eliminates this spurious dependence but in fact increases the chance of correctly determining the hierarchy.

\vfill

\begin{flushleft}
{\today}
\end{flushleft}
\end{titlepage}

\hfill{}


\setcounter{footnote}{0}

\noindent
This year the Daya Bay \cite{dayabay,neut2012} and RENO \cite{reno} experiments have demonstrated beyond any reasonable doubt that $\theta_{13}$ is as much as an order of magnitude larger than had been suspected several years ago.  
This large value of $\theta_{13}$ implies that 1-3 neutrino oscillations may be observed at medium baselines, which we define to be 40-80 km.  The medium baseline neutrino spectrum may then be used to determine the neutrino mass hierarchy \cite{petcovidea}.  Such experiments are now not only practical but indeed they will be performed within the next decade \cite{caojunseminario,renonuturn,yifangseminario}.  

However at these baselines, due to a degeneracy in the high energy neutrino spectrum \cite{parke2007,oggi,noi}, a determination of the hierarchy requires a measurement of 1-3 oscillations at low neutrino energies $E$.  As a result of the finite energy resolution of the detector and various interference effects \cite{noi} these low energy peaks are difficult to identify individually at medium baselines $L$.  Nonetheless if the nonlinear energy response of a detector is well understood then one may measure {\it{the sum of the peaks}} by studying the $k\sim\mn31/2$ region of the  $L/E$-Fourier transform of the neutrino spectrum \cite{hawaii}.  The most popular variables for such a determination are the fractional difference $RL$ between the deepest minima of the Fourier cosine transform and the difference $PV$ between the deepest minimum and the highest peak of the Fourier sine transform \cite{caojun,caojun2}.  Although these two variables are somewhat degenerate, an improvement may be obtained by considering their sum $RL+PV$.

A serious obstruction to this analysis, and thus to plans to measure the neutrino mass hierarchy at medium baselines, has been described in Ref.~\cite{oggi}.  The authors observed that the combination $RL+PV$ is very sensitive to the choice of model of the reactor neutrino flux $\Phi(E)$ and to variations of $\mn32$ which are smaller than the precision to which this mass difference has been determined by MINOS \cite{minos}.  While the observed shift appears to depend upon both $\mn32$ and the hierarchy, from Fig. 4 of Ref.~\cite{oggi} it can be seen that the shift depends only upon the effective mass difference \cite{parke2005,noi}
\beq
\meff=\cp212\mn31+\sp212\mn32.
\eeq
The neutrino flux from reactors is known poorly.  The theoretical normalization has recently increased by about 3\% \cite{nuovoflusso} and the 6+ MeV flux has increased by an additional 3\% \cite{huber}.  The flux beyond about 8 MeV is not known at all due to its strong dependence upon decays of exotic isotopes \cite{nuovoflusso}.  Even worse, all of these theoretical fluxes are about 6\% above the observed fluxes at very short \cite{reattoreanom} and 1 km \cite{noiunokm} baselines.  Thus the large sensitivity of $RL+PV$ upon the poorly known fluxes and $\meff$  appreciably reduces the probability that a medium baseline reactor experiment can correctly determine the neutrino mass hierarchy.

We will now explain the cause of this strong dependence.  As the dependence of $RL$, $PV$ and $RL+PV$ upon these parameters is virtually indistinguishable, for brevity we will consider only 
\beq
RL=\frac{R-L}{R+L}
\eeq
which is the fractional difference between two minima $R$ and $L$ of the Fourier cosine transform of the neutrino spectrum 
\beq
F_c(k)=\int d\left(\frac{L}{E}\right)\frac{E^2}{L} \frac{\Phi(E)\sigma(E)}{4\pi L^2}  P_{ee}\left(\frac{L}{E}\right) \cos\left(\frac{kL}{E}\right) \label{fc}
\eeq
where the tree level neutrino inverse $\beta$ decay cross section is \cite{sezionedurto}
\beq
\sigma(E)=0.0952\times 10^{-42}\mathrm{cm}^2\frac{E_e\sqrt{E_e^2-m_e^2}}{\mathrm{MeV}^2}\hsp
E_e=E-m_n+m_p
\eeq
and the electron neutrino survival probability is
\bea
P_{ee}&=&\sp413+\cp412\cp413+\sp412\cp413+\frac{1}{2}(P_{12}+P_{13}+P_ {23})\nonumber\\
P_{12}&=&\spp2212\cp413\cos\left(\frac{\m21L}{2E}\right)\hsp
P_{13}=\cp212\spp2213\cos\left(\frac{\mn31L}{2E}\right)\nonumber\\
P_{23}&=&\sp212\spp2213\cos\left(\frac{\mn32L}{2E}\right).\nonumber
\eea
Following Ref.~\cite{caojun}, a
 $3\%/\sqrt{E}$ energy resolution is included by convoluting the observed energy spectrum with
\beq
\rm{exp}\left(-\frac{(E-E\p)^2}{0.0018(E_e+m_e)MeV}\right).
\eeq
We use the neutrino mass matrix parameters of Ref.~\cite{caojunseminario}.

As was demonstrated in Ref.~\cite{noi}, the minima whose difference defines $RL$ lie just on either side of $k=\mn31/2$.  These minima arise from the Fourier transform of $P_{13}$ which is independent of the hierarchy, but the contribution of $P_{23}$  provides a perturbation which makes the right (left) minimum deeper for the normal (inverted) hierarchy.

\begin{figure} 
\begin{center}
\includegraphics[width=5.2in,height=2in]{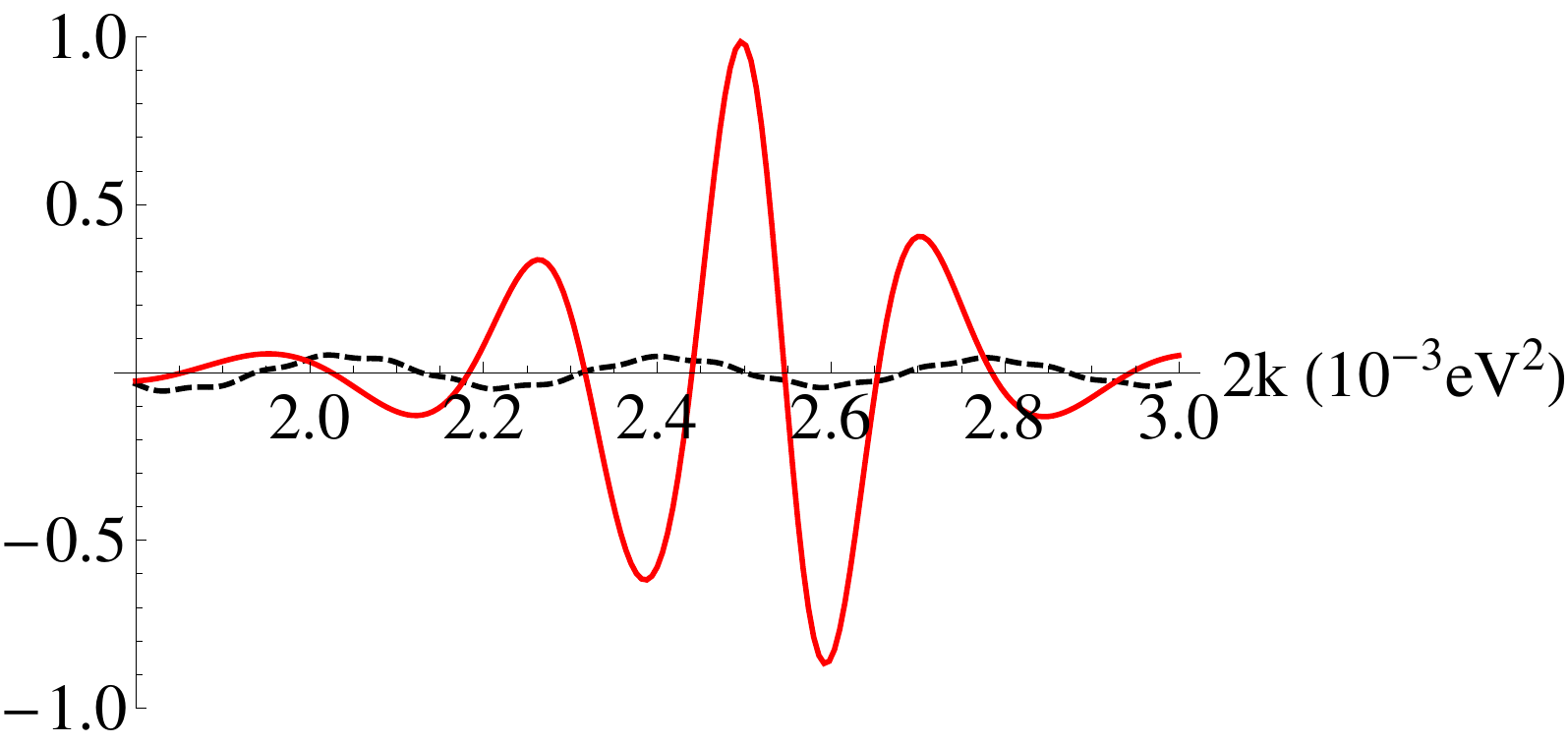}
\caption{The cosine transforms of the unoscillated flux (black dashed curve) and the full $P_{13}+P_{23}$ oscillated flux (red solid curve) are shown.  As $\meff$ varies, the $P_{13}+P_{23}$ peaks move and so the interference between the two contributions to the cosine transform of the neutrino spectrum varies.  The reactor fluxes used are those of the 1980's.}
\label{vecflufig}
\end{center}
\end{figure}

The problem observed in Ref.~\cite{oggi} is that, depending upon the reactor flux model used, the transform of the unoscillated reactor flux $\Phi(E)\sigma(E)E^2/L^3$ itself may contribute peaks near $k=\mn31/2$ which interfere with those of $P_{13}+P_{23}$ and so affect $RL$.  While the cosine transform of the unoscillated flux $\Phi(E)\sigma(E)E^2/L^3$ is independent of the neutrino mass splittings, the locations of the peaks of $P_{13}+P_{23}$ are proportional to $\meff$.  This means that the relative phase between the Fourier transform of the unoscillated spectrum and that of $P_{13}+P_{23}$ depends on the precise value of $\meff$.   As a result the oscillations in the Fourier transform of $\Phi(E)\sigma(E)$ lead to an $\meff$-dependence in the quantity $RL$ just of the kind observed in Ref.~\cite{oggi} using old reactor flux models.  In fact, using the $\u35$ flux from Ref.~\cite{uflusso}, the $\pu39$ and $\pu41$ fluxes from \cite{pflusso} and the Gaussian approximated $\u38$ flux from Ref.~\cite{quadflusso} with the isotope ratios of Ref.~\cite{caojun} we find an oscillation in the unoscillated spectrum term in Eq.~(\ref{fc}).  Using this old model of the reactor flux, in Fig.~\ref{vecflufig} we compare the Fourier transform of the unoscillated term with that of the $P_{13}+P_{23}$ term, which is sensitive to the hierarchy.  One can see that the unoscillated term is periodic with the same wavelength as was observed in Fig. 4 of Ref.~\cite{oggi}, and thus the interference between these two terms oscillates as $\meff$ varies, shifting the $P_{13}+P_{23}$ peaks and so reproducing the effect reported in that note.

\begin{figure} 
\begin{center}
\includegraphics[width=5.2in,height=2in]{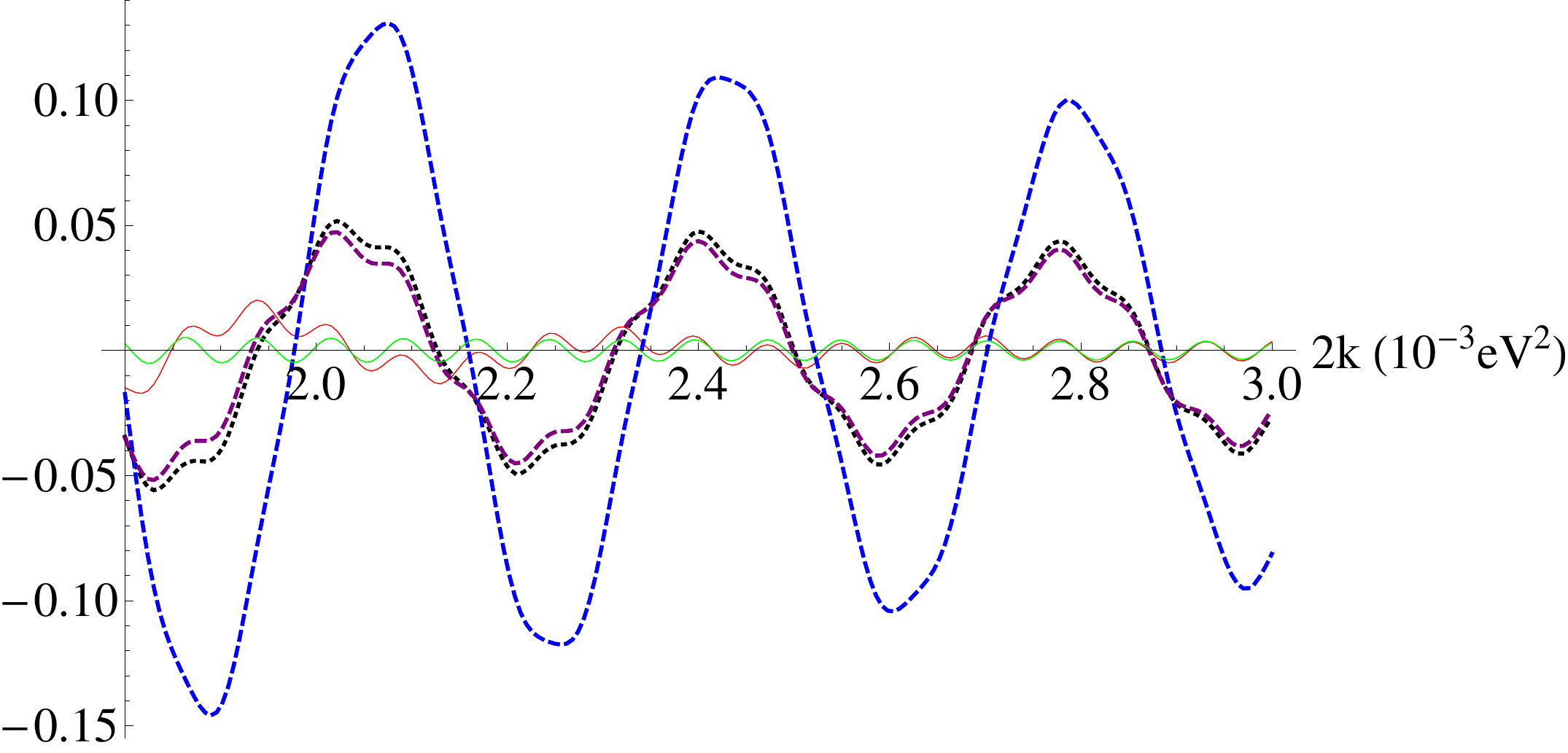}
\caption{The cosine transforms of the unoscillated flux is shown for numerically interpolated fluxes from the 1980's \cite{uflusso,pflusso,quadflusso} (black dotted curve), for a quadratic fit to fluxes  from the 1980's \cite{quadflusso} and for quintic fits of the new fluxes of Ref.~\cite{huber}.  The latter two are shown with cutoffs of 8.5 MeV (dashed curves) and 12.8 MeV (solid curves).  The blue dashed curve corresponds to the quadratic fit flux.  The red and green solid curves, corresponding to 12.8 MeV cutoffs, are close to zero.  Therefore the interference effect is present if the cutoff is at 8.5 MeV and but not if the fits are naively extrapolated to 12.8 MeV.  This demonstrates that $RL$ is sensitive to the neutrino spectrum above 8.5 MeV.}
\label{cutofffig}
\end{center}
\end{figure}

Ref.~\cite{oggi} concludes that this strong dependence of $RL$ upon the reactor flux means that a precise knowledge of this flux is desirable to determine the neutrino mass hierarchy at a short baseline experiment.  Our conclusions differ.  This difference arises from the observation that, as can be seen in Fig.~\ref{cutofffig}, near $k=\mn31/2$ the oscillations of the cosine transform of the reactor spectrum are governed by the highest energy neutrinos (8+ MeV) which, according to Refs.~\cite{parke2007,noi}, are not sensitive to the hierarchy.   On the other hand in Fig.~\ref{cutofffig} we see that the effect is present in the case of both old and new fluxes if the spectrum is cut off at 8.5 MeV while it is negligible if the spectrum is cut off at 12.8 MeV.   However since the quadratic and quintic fits to these fluxes are only reliable below 6.5 MeV and marginally reliable below 8 MeV, there is no reason to trust a naive extrapolation to 12.8 MeV.

{\it{As $RL$ depends strongly on the spectrum between 8.5 and 12.8 MeV, which in turn is independent of the hierarchy, this high energy spectrum provides a nuisance parameter for the determination of the hierarchy using $RL$.}}  The solution suggested in Ref.~\cite{oggi} is to determine the spectrum precisely, however so few neutrinos are observed in this range that such a determination would be difficult, indeed the spectrum is not understood at the required precision even at the energies with high fluxes \cite{reattoreanom}.  Even if such a measurement were possible, then $RL$ would still be likely to depend upon $\meff$ with a higher sensitivity than the mass determination at MINOS and probably at NO$\nu$A, making a determination of the hierarchy at a medium baseline more challenging.

\begin{figure} 
\begin{center}
\includegraphics[width=5.2in,height=2in]{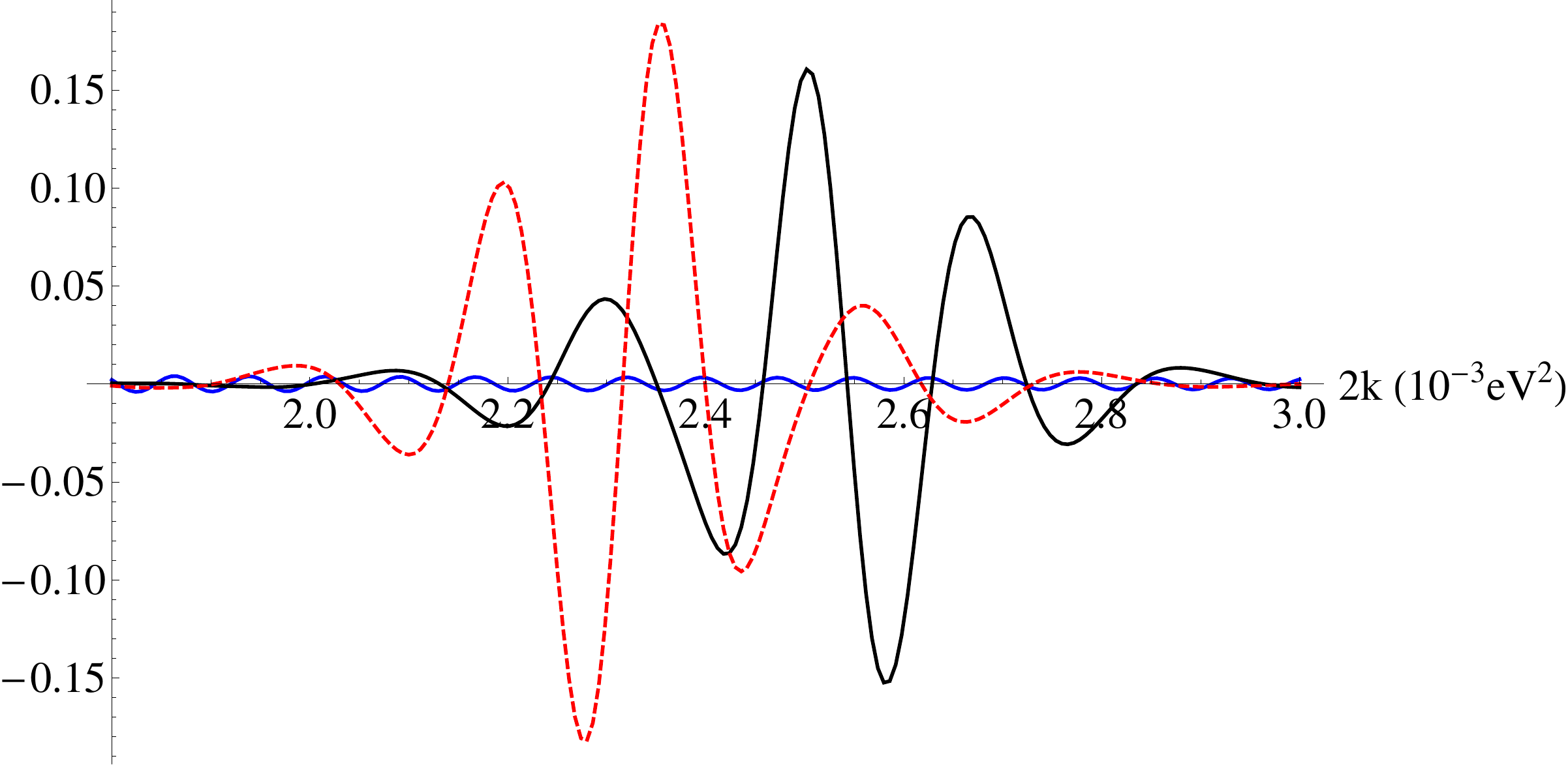}
\caption{Here we see the weighted Fourier transform of the spectrum without oscillations (blue solid curve) and with oscillations in the case of the normal (black solid curve) and inverted (red dashed curve) hierarchies.  One can see that the solid, blue unoscillated curve is very close to zero.  We have checked that this curve is essentially independent of the cutoff and so the reactor spectrum no longer affects $RL$.   Comparing with Fig.~\ref{vecflufig} one can see that the difference $RL$ between the depths of the minima is even greater in this weighted case, allowing for a better determination of the hierarchy than was possible with an unweighted Fourier transform.}
\label{xformpesfig}
\end{center}
\end{figure}

Our solution is to replace $RL$ and $PV$ with indicators that are insensitive to the high energy neutrino spectrum, by providing an energy-dependent weight $w(E)$ on the neutrino spectrum in the Fourier transform.  As we saw in Fig.~\ref{cutofffig}, a simple cutoff in the Fourier transform will amplify the spurious dependence.  The weight needs to cut off the high energies gradually, with derivative scales much longer than $\mn31$, so as to not itself introduce spurious peaks in the critical part of the Fourier transforms.  One such choice of weight which we have found works quite well is a Gaussian
\beq
F_c(k)=\int d\left(\frac{L}{E}\right)e^{-\frac{0.08E^2}{\rm{MeV}^2}}\frac{E^2}{L} \frac{\Phi(E)\sigma(E)}{4\pi L^2}  P_{ee}\left(\frac{L}{E}\right) \cos\left(\frac{kL}{E}\right).
\eeq
The same weight serves well in both the sine transform and also the nonlinear transforms of Ref.~\cite{noi} which determine the hierarchy more reliably than $RL+PV$ at baselines below about 55 km \cite{noisim}.  As can be seen by comparing the unweighted and weighted cosine transforms in Figs.~\ref{vecflufig} and~\ref{xformpesfig}, not only does the weighting procedure preserve $RL$, but it actually increases the difference in the peak sizes between the normal and inverted hierarchies.  Thus this solution to the dependence upon the high energy neutrino tail not only removes the spurious dependence, for any high energy reactor spectrum, but it increases the chance of success of the determination of the hierarchy.  In simulations \cite{noisim} we will show that when the weight function is optimized with a neural network this increase is of order~2\%.

\section* {Acknowledgement}

\noindent
JE is supported by the Chinese Academy of Sciences
Fellowship for Young International Scientists grant number
2010Y2JA01. EC and XZ are supported in part by the NSF of
China.  


\end{document}

\section{Introduction}

\section{The Monte Carlo}

\subsection{The Parameters}

This paper builds upon the simulations of medium baseline neutrino oscillation in Ref. \cite{caojun2}.  For ease of comparing the results of that study and ours, we have adopted many of the same for the neutrino mass matrix
\beq
\m21=7.6\times 10^{-5}\rm{\ eV}^2\hsp
\mn22=2.4\times 10^{-3}\rm{\ eV}^2\hsp
\sp212=0.32
\eeq
and for the absolute detector energy resolution
\beq
\sigma_E=.03\sqrt{E_ {pr}\rm{(MeV)}}
\eeq
where the prompt energy $E_{pr}$ is related to the neutrino energy $E$ and the positron energy $E_e$ by
\beq
E_{pr}=E_e+m_e=E-m_n+m_p+m_e\sim E-780\ \mathrm{keV}\hsp
\eeq

While these parameters have changed very little over the past 3 years, the motivation for our analysis is the large increase in $\theta_{13}$ \cite{dayabay,reno,doublechooz}.  We will use Daya Bay's most recent result \cite{neut2012}
\beq
\spp2213=0.089.
\eeq 
This increase is critical to the determination of the hierarchy at a neutrino disappearance experiment, as the hierarchy manifests itself via energy-dependent shifts of the 1-3 oscillation peaks \cite{petcovidea,parke2007}.

\subsection{The Simulation}

The overall reactor flux normalization is fixed to be that of Ref \cite{caojun2}.  It is asserted that the Daya Bay plus Ling Ao I and II reactor complexes, which together have 17.4 GW of thermal capacity, lead to the observation of 25,000 neutrinos at a 20 kton detector with a baseline of 60 km.  To be precise this does not fix the reactor flux normalization, but rather the overall normalization of the product of the reactor flux, the antineutrino cross section of the target, the detector efficiency and the effect of neutrino oscillations.  This condition, together with the tree level inverse $\beta$ decay cross section in Ref. \cite{sezionedurto}
\beq
\sigma(E)=0.0952\times 10^{-42}\mathrm{cm}^2(E_e\sqrt{E_e^2-m_e^2}/\mathrm{MeV}^2)
\eeq
and the oscillation probability $P_{ee}$
\bea
P_{ee}&=&\sp413+\cp412\cp413+\sp412\cp413+\frac{1}{2}(P_{12}+P_{13}+P_ {23})\nonumber\\
P_{12}&=&\spp2212\cp413\cos\left(\frac{\m21L}{2E}\right)\hsp
P_{13}=\cp212\spp2213\cos\left(\frac{\mn31L}{2E}\right)\nonumber\\
P_{23}&=&\sp212\spp2213\cos\left(\frac{\mn32L}{2E}\right)\nonumber
\eea
are then used to normalize the effective reactor flux $\Phi(E)$ per time per GW of thermal capacity.  This reactor flux is effective in the sense that it is already multiplied by the efficiency of the detector.

With this normalization in hand, the average number density of antineutrinos at energy $E$ from a given reactor which without oscillation would be observed during 3 years at a 20 kton detector at a baseline $L$ is then
\beq
N(E)P_{ee}(L/E)=\frac{\Phi(E)\sigma(E)P_{ee}(L/E)}{4\pi L^2}.
\eeq
In each 60 kton year experiment we simulate precisely $N(E)$ neutrinos per unit energy from each reactor.  We will recreate 120 and 300 kton year experiments by simply summing the neutrinos from pairs or quintuplets of 60 kton year experiments.  To minimize the relative statistical errors between experiments with different exposure times,120 and 300 kton year experiments will be created from 60 kton experiments reported on the same tables.


Unlike Ref. \cite{caojun2,caojunseminario} our total neutrino flux in each experiment depends on $L$.  In fact it increases faster than $1/L^2$ as $L$ is decreased from 58 km because there is less loss due to 1-2 neutrino oscillations.  This means that at short baselines, such as 40-50 km, our simulations allow a precise determination of the high energy peaks, which are hardly depleted by 1-2 oscillation.  The energies of these peaks $2\pi/\Delta M^2_{\rm{eff}}$ determine the effective mass splitting \cite{parke2005,noi}
\beq
\Delta M^2_{\rm{eff}}=\cp212\mn31+\sp212\mn32. \label{meff}
\eeq
As a result we will see that our simulations favor shorter baselines than those preferred in Refs~\cite{caojun2,caojunseminario}.  We will also see that shorter baselines are preferred as they reduce the fractional backgrounds from distant reactors, backgrounds not included in previous studies.

Once we have fixed the numbers and energies of the neutrinos arriving at our detector, the finite resolution $\sigma_E$ of the detector is applied.   With infinite luminosity the observed neutrino spectrum $P_{obs}(E)$ would be the theoretical spectrum convoluted with a Gaussian of width $\sigma_E$.  To implement this effect at a finite luminosity we shift the energy of a neutrino by a random variable  $\delta$ with Gaussian probability density $\rm{exp}(-\delta^2/2\sigma_E^2)/\sqrt{2\pi\sigma_E}$.  

Thus our simulation correctly accounts for statistical errors in the determination of the energy due to the finite number of photoelectrons detected.  However it does not take into to account for a systematic nonlinear error in the determination of energy.  While linear errors are harmless to the Fourier analysis that will be performed later, they simply shift the points, even a small linear uncertainty can alter \cite{petcov2010} or destroy \cite{oggi} the effectiveness of such an analysis.  In practice the energy can be calibrated, at least at some energies and at some points inside of the detector, by inserting radioactive sources.  A peak by peak analysis of the spectrum can be performed just using peaks at these energies \cite{noi} but this requires more flux than a Fourier analysis.

\subsection{Improvements over Previous Simulations}

Although the simulation does not account for the unknown nonlinear response of the detector, it does include an interference effect \cite{noi} which has so far not been simulated in the literature.  The reactor complex at Daya Bay is separated by more than a kilometer from the reactor complexes Ling Ao I and Ling Ao II.  While this set of reactors is in general an ideal choice for medium baseline experiments, given the high flux and the existence of 8 short baseline detectors, in general the 1-3 oscillations from Ling Ao and Daya Bay will be out of phase at low energies.  The measurement of 1-3 oscillations at these low energies are essential to break a degeneracy which prohibits the determination of the neutrino mass hierarchy \cite{parke2007,oggi,noi}.  Thus such interference poses a severe problem for the determination of the hierarchy.  However, since the Daya Bay, Ling Ao I and Ling Ao II complexes, like RENO, lie along a line, it is possible to evade this interference effect if the detector is roughly orthogonal to the line.  The price of this choice is that one cannot benefit from the flux at the proposed HuiDong reactor complex, which would no longer be equidistant from the detector.  

Our simulation also includes the flux from many distant reactors, of which TaiShan and YangJiang will begin to come online next year.  We will see that these backgrounds have a nonnegligible effect on the probability of determination of the neutrino mass hierarchy.   They provide both more backgrounds but also potentially more usable information for the determination of $\theta_{12}$.

\section{Data Analysis}

\subsection{Fourier Analysis and Reactor Flux Models} \label{flusssez}

In this note we will not consider a $\chi^2$ or peak energy analysis of the data, although these require an understanding of the nonlinear response of the detector only in limited regions.  Instead we will consider a Fourier analysis \cite{hawaii}, which has the advantage that it combines neutrinos from multiple peaks and so requires less flux, despite the fact that it requires an unprecedented understanding of the nonlinear response of the detector over a wide range of energies. Following Refs.~\cite{caojun,caojun2} we will decompose this transform into a real and a complex part
\bea
F_c(k)&=&\int d\left(\frac{L}{E}\right)\frac{E^2}{L} N(E)  P_{ee}\left(\frac{L}{E}\right) \cos\left(\frac{kL}{E}\right)\nonumber\\
F_s(k)&=&\int d\left(\frac{L}{E}\right)\frac{E^2}{L} N(E)  P_{ee}\left(\frac{L}{E}\right) \sin\left(\frac{kL}{E}\right) \label{fcos}
\eea
where $N(E)$ is the number of neutrinos in the energy bin centered at energy $E$ in a given experiment and the factor $E^2/L$ is the Jacobian factor which transforms a neutrino density in $E$ into a density in $L/E$.

We will be interested in the form of the Fourier transforms near the 1-3 oscillation peak $k=\mn31/2$.  At these wavenumbers it is commonly believed that gross features of the spectrum are independent of the reactor flux and $P_{12}$, but this is not quite true.  Recently Ref.~\cite{oggi} found that the quantities (\ref{rlpv}) used in the analysis of Refs.~\cite{caojun,caojun2} depend strongly upon the reactor flux model used and for some flux models even upon $\mn32$, with variations as large as a factor of 5.  The authors concluded that a precise determination of the reactor spectrum is necessary to determine the hierarchy.  This is also not quite true.  

Instead the effect discovered in Ref. \cite{oggi} is the strong dependence of the quantities (\ref{rlpv}) upon the high energy part of the neutrino flux spectrum, in particular above about 8 MeV.  Even though the neutrino flux at such high energies is very small, it can be leading contribution to the Fourier transform of the unoscillated neutrino flux $\Phi(E)$ at $k\sim\mn31/2$.  For example a cutoff even as high as 8.5 MeV can lead to fluctuations in the Fourier transform of the reactor flux $\Phi(E)$ which are as large as the fluctuations in $P_{13}+P_{23}$ which determine the quantities (\ref{rlpv}) and therefore the hierarchy.  These oscillations in the Fourier transform of $\Phi(E)$ in the presence of a cutoff have just the same wavelength as those reported in Ref. \cite{oggi}.  The $\mn32$-dependence observed in Fig. 4 of that reference arises from the $\meff$ dependence of the $P_{13}+P_{23}$ peaks, which determines the interference with the $\mn32$-independent transform of the reactor flux.  This poses a real problem as the $\mn32$-dependence is large on scales within the MINOS error on the determination of $\mn32$, and so it cannot be simply corrected during the data analysis.

How can this problem be resolved?  Ref. \cite{oggi} suggests an accurate reactor flux determination, and to back up this claim shows that the effect is minimized using the new fluxes of Refs. \cite{reattoreanom,huber2011}.  However this approach only appeared to work because the fit to the neutrino flux of Refs. \cite{reattoreanom,huber2011} is of a form such that, when extrapolated to high energies, the contribution to the $k\sim\mn31/2$ oscillations is small.  This is not because those fluxes are correct at such high energies, indeed above 6.5 MeV they begin to become unreliable as multistage decays of very excited isotopes become important \cite{reattoreanom}.  It is simply a feature of the extrapolation of a particular fitting function beyond its range of validity.  In fact the Gaussian fitting functions used in Refs. \cite{caojun,caojun2,noi} can be extrapolated to arbitrarily high energy and give no discernible contribution at all near $k=\mn31/2$.  

The real high energy neutrino spectrum at these high energies is unknown, but there is no reason to believe that it will be close to a function of a special form with a compactly supported Fourier transform.   Thus even if at some future date the high energy reactor spectrum can be measured precisely, the strong $\mn32$-dependence in the Fourier transform quantities will still be present and so it will not be useful in the determination of the hierarchy.  

In fact the general arguments in Refs. \cite{parke2007,noi} make it obvious that the high energy spectrum cannot contribute to a determination of the hierarchy.  Neutrinos at such high energies have not oscillated enough times to depend observably on the hierarchy.  Thus we conclude that an accurate determination of the high energy neutrino spectrum, in contradiction with the claims in Ref. \cite{oggi}, is unlikely to resolve to strong $\mn32$-dependence which they observed reduces the chance of successfully determining the hierarchy.

This problem greatly reduces the probability of success of a determination of the hierarchy using the variables (\ref{rlpv}), but it does not affect the positions of the peaks, which after all contain all of the information concerning the hierarchy.  The problem is created by a contribution to the analysis parameters (\ref{rlpv}) by neutrinos whose energy is so high that they are not sensitive to the hierarchy.  This is similar to the contribution of unoscillated high energy neutrinos to RENO's determination of $\theta_{13}$ \cite{reno}.  Just as that problem was solved in Ref. \cite{renofrancia} by a cutoff eliminating the spurious signal from high energy neutrinos, here it may be solved by an energy-dependent weight in the Fourier transform which reduces the impact of the high energy part of the spectrum.  We have checked that this can be arranged by simply inserting a scale factor in the Fourier transform, such as exp$(-0.08 E^2/{\rm{MeV}}^2)$, which suppresses the high energies near the cutoff by more than the lower energies which are sensitive to the hierarchy.  

\subsection{Observables}

Now that we have inserted a scale factor in the Fourier transform, the qualitative features of the transformed spectrum depend primarily upon $P_{13}$, which gives a symmetric $F_c(k)$ with a central peak and damped oscillations with a wavenumber of order $L\langle 1/E\rangle$ where $\langle 1/E\rangle$ is the average value of $1/E$.  At medium baselines, of order 40-80 km, this wavenumber is of order $\m21$.  Similarly the $P_{13}$ contribution to the sine transform $F_c(k)$ is antisymmetric about $k=\mn31/2$, with a maximum at higher $k$ and a minimum at lower again.  Again, as one varies $k$ from $k=\mn31/2$ one finds a series of ever shrinking oscillations separated by a characteristic distance of order $\m21$.  At $L=58$\ km analytic approximations to these features are derived in Ref.~\cite{noi}.  As the Fourier transforms are linear, they add the transform of the $P_{23}$ oscillations to that of the larger $P_{13}$ oscillations.

The hierarchy can be determined from the way in which the resulting asymmetric perturbation.  It was observed in Ref.~\cite{caojun} and derived in Ref.~\cite{noi} that in the case of the normal (inverted) hierarchy the $P_{23}$ oscillations render the first minimum $R$ to the right of the global maximum of $F_c$ larger (smaller) than its mirror image $L$ and similarly render the maximum $P$ of $F_{s}$ just to the right of $k=\mn31/2$ larger (smaller) than the minimum just to its left $V$.    Ref~\cite{caojun} introduced two parameters which characterize these effects
\beq
p_1=\frac{R-L}{R+L}\hsp
p_2=\frac{P-V}{P+V} \label{rlpv}
\eeq
finding that positive (negative) values of these two parameters tend to indicate the normal (inverted) hierarchy.

In Ref.~\cite{noi} two more parameters were suggested.  First, the value $\phi$ of $P_s$ at the maximum of $P_c$, which is positive (negative) for the normal (inverted) hierarchy.  Second, the difference in values $m_\pm$ of the maxima of hierarchy-dependent nonlinear Fourier transforms
\beq
F_n^\pm(k)=\int d\left(\frac{L}{E}\right)\frac{E^2}{L}\Phi(L/E) P_{ee}(L/E) g(L/E) \cos\left(k\frac{L}{E}\pm 2\pi\alpha\left(\frac{k}{2\pi}\frac{L}{E}\right)\right).
\eeq
These can be encoded into the normalized observables
\beq
p_3=\frac{\phi}{P+V}\hsp
p_4=\frac{m_+-m_-}{m_-+m_+}.
\eeq

In our analysis we have also introduced one new parameter.  We observed that statistical errors in Fourier space tend to have correlation lengths which are longer that the $\m21$-scale wavelength of oscillations in the Fourier transform of $P_{13}+P_{23}$.  Thus if one peak is erroneously too low or too high, its nearest neighbors are also likely to be too low or too high.  It was shown in Ref. \cite{noi} that as $k$ varies from the center point $\mn31/2$, the analytic (no statistical error) heights of the peaks shrinks exponentially, so the secondary peaks and valleys are much smaller than the primary peaks and valleys, which we define to be those closest to the center.  

The observables above all depend upon the heights of the primary peaks and valleys.  But this correlation means that when a primary peaks or valleys is too high (low) due to a statistical error, the corresponding secondary peak or valley should also be too high (low).  Thus one can reduce the statistical errors by using the size of the secondary oscillations to correct those of the primary oscillations.  This can be implemented automatically by defining a secondary analog of $p_1$
\beq
p_5=\frac{L_2-R_2}{R+L}
\eeq
where $R_2$ and $L_2$ are the depths of the secondary valleys, which are defined to be the minima within a distance $2\m21$ to the right and left of the corresponding primary valleys respectively.  Thus an indicator which is a linear combination of $p_1$ and $p_5$ can be expected to be less susceptible to statistical errors than $p_1$ alone.  Of course secondary analogs of the other indicators can also be introduced, but we expect them to be largely degenerate with $p_5$.

\subsection{Neural Network}

In the previous subsection we identified five quantities $p_i$ which may be used to determine the hierarchy.  The sign of any one of the first four is already a good indicator of the hierarchy, in fact they are quite degenerate.  The fifth indicator $p_5$ alone is an inferior indicator of the hierarchy, but it characterizes the direction of the statistical errors of the others.

Our goal is to determine the best indicator, that which, given a dataset, has the highest probability of correctly determining the mass hierarchy.  As these indicators are not precisely degenerate, the best indicator will not be a single $p_i$ but rather some combination of all 5.  For simplicity we will consider linear combinations
\beq
I=c_0+\sum_{i=1}^{5}c_i p_i \label{secondo}
\eeq 
where $c_i$ are 6  constants.  The constants will be chosen such that, given a set of experiments in which the neutrino mass hierarchy is normal in as many as it is inverted, the chance of success is the highest.  The chance of success is defined to be the average of the percentage of normal hierarchy experiments for which $I$ is positive with the number of inverted hierarchy experiments for which $I$ is negative.  Clearly the overall scale of $c_i$ is irrelevant, but the optimal choice of $c_i$ in general depends on the baseline, the reactors that are operational, the geometry of the reactors with respect to the detector and even the time for which the experiment has run.

We will optimize $c_i$ given a dataset, in general of 500 experiments of each hierarchy, by beginning at $c_i=0$ with a fixed temperature, and then at each step both cooling and moving the vector $c_i$ a temperature-dependent distance in a random direction.  If the probability of success is improves, the new location is the starting point for the next recursion.  After a fixed number of steps the current value of $c_i$ will be fixed.  It will then be applied to another set of 500 experiments of each hierarchy to determine the chance of success.  It is important that a separate set of experiments is used for this determination to avoid overfitting the data, which would lead to an artificially inflated chance of success.  Then the best $c_i$ will be found with the second set of data and the result applied to the first, to obtain a second chance of success.  The chance of success that we will report will be the average of these two numbers.

In general we have found that this 1-layer neural network method leads to an improvement over the $p_1+p_2$ quantity in Ref. \cite{caojun} of 1-2 percent.  This can be increased yet further with a second layer, applied before first layer.  We have seen in Subsec. \ref{flusssez} that in order to avoid the spurious dependence of $p_i$ upon the high energy neutrino spectrum observed in Ref. \cite{oggi} one may include energy dependent weights $w(E)$ in the functionals
\bea
\tilde{F}_c[w(E)](k)&=&\int d\left(\frac{L}{E}\right)w(E)\frac{E^2}{L} N(E)  P_{ee}\left(\frac{L}{E}\right) \cos\left(\frac{kL}{E}\right)\nonumber\\
\tilde{F}_s[w(E)(k)&=&\int d\left(\frac{L}{E}\right)w(E)\frac{E^2}{L} N(E)  P_{ee}\left(\frac{L}{E}\right) \sin\left(\frac{kL}{E}\right).
\eea
Here $E$ is the observed energy, which due to the finite energy resolution of the detector is not precisely equal to the true neutrino energy.

We noted that $w(E)=exp(0.08 E^2/\rm{MeV}^2)$ removes the spurious high energy dependence of Ref. \cite{caojun}.  What about other choices of weights?  In principle, for each Fourier transform one can consider weights to be an function of some arbitrary parameters.  One can then try to optimize these parameters together with the $c_i$, but optimizing so many parameters at once is not possible with the small number of experiments that we have generated.  However, in a first layer of the neural network, for each $i$, except for $i=5$ as $p_5$ does not determine the hierarchy well on its own, one can optimize the parameters in the weight function so as to maximize the chance of success of that $p_i$.  Then one can optimize the $c_i$ for these new and improved $p_i$.

For example, one may choose the weight functions $w(E)$ so as to deemphasize the spurious high energy neutrinos, to accentuate the neutrinos around the 1-2 oscillation maximum or to accentuate the lowest energy neutrinos whose signals are the most strongly hierarchy-dependent.  In realistic detectors in which the nonlinear response is only known in some region, where it has been calibrated by radioactive decays, one may wish to weight those regions more heavily.  This can be done, for example, by using a weight function which contains a Gaussian distribution centered upon the energy range to be accentuated.   The only constraint on the weight function is that its characteristic derivative scale be smaller than $\mn31$, so that it does not itself produce a spurious contribution to the Fourier transform near $k=\mn31/2$ and so to the quantities $p_i$.

One such choice of weight coefficients for the cosine transform is
\beq
F_c\p(k)=\tilde{F}_c[1](k)+a_1\tilde{F}_c[e^{-E/(3\rm{\ MeV})}](k)+a_2\tilde{F}_c[e^{-E^2/(12\rm{\ MeV}^2)}](k). \label{primo}
\eeq
Here the first weighted transform $\tilde{F}_c[1]$ is just the unweighted transform defined in Eq. (\ref{fcos}) while the second weights lower energy neutrinos more heavily and the third is that of Subsec. \ref{flusssez}.  A given $p_i$ can then be computed with the new $F\p$ and the weight coefficients $a_j$ can be optimized for each value of $i$ separately.

Summarizing, for each value of $i$ from 1 to 4 the first layer of the neural network optimizes $a_1$ and $a_2$ of Eq. (\ref{primo}) so as to maximize the probability of success of $p_i$ alone.  Then the second layer optimizes the $c_i$ in Eq. (\ref{secondo}) so as to maximize the probability of success of the indicator $I$.

\section{Optimizing the Baseline and Geometry}

\subsection{The Optimal Baseline}

\begin{table}[position specifier]
\centering
\begin{tabular}{c|l|l|l}
Baseline&$N_{\overline{\nu}_e}$ at 60 ky&Hierarchy at 60 ky&Hierarchy at 120 ky\\
\hline\hline
42 km&64,860&86.5\%\ (80.8\%)&\%\ (91.4\%)\\
\hline
44 km&55,578&86.6\%\ (82.6\%)&\%\ (91.2\%)\\
\hline
46 km&48,030&87.0\%\ (82.3\%)&\%\ (91.6\%)\\
\hline
48 km&41,900&87.8\%\ (86.0\%)&\%\ (94.2\%)\\
\hline
50 km&36,931&88.2\%\ (86.7\%)&\%\ (95\%)\\
\hline
52 km&32,915&88.0\%\ (87.1\%)&\%\ (94\%)\\
\hline
54 km&29,679&87.9\%\ (87.4\%)&\%\ (95\%)\\
\hline
56 km&27,080&\%\ (83.9\%)&\%\ (94\%)\\
\hline
58 km&25,000&\%\ (80.6\%)&\%\ (92\%)\\
\hline
60 km&23,342&83.0\%\ (82.3\%)&\%\ (93\%)\\
\hline
\end{tabular}
\caption{The observed $\overline{\nu}_e$ flux from a single 17.4 GW reactor and the probability of determining the hierarchy as a function of baseline and number of kton years determined using the neural network.  The neural networks trained on the neighboring baselines, and the values of the coefficients $c_i$ used to determine the chance of success are the averages of the coefficients at the available neighboring baselines.  Probabilities in parenthesis are those using $p_1+p_2$ as in Ref.~\cite{caojun}.}
\label{disttab}
\end{table}

As a first application of our simulation we have attempted to determine the optimal baseline in an idealized situation in which all 17.4 GW of thermal capacity are located at a single reactor, at a fixed baseline.  This eliminates the interference effects of Ref. \cite{noi} which will be investigated in Subsec. \ref{intsez}.  The probability of success then depends only upon the baseline, the number of ktons of the detector multiplied by the number of years of observation and the method of data analysis.  We will consider a Fourier transform analysis with two indicators, $p_1+p_2$ which was introduced in Ref. \cite{caojun} and also our neural network.  The results are displayed in Table \ref{disttab}.

\subsection{Interference Between Reactors in the Same Complex} \label{intsez}

No single reactor produces enough flux to determine the neutrino hierarchy in a reasonable amount of time.  Thus it is inevitable that complex of reactors need to be used, and often the distances between these reactors is considerable.  For example in Ref. \cite{weihai} the proposed cite for Daya Bay II is 3.5 km closer to the reactor complexes at Daya Bay and Ling Ao than to that planned at HuiDong.  These different baselines mean that for some energies the neutrinos from one reactor will arrive at the 1-3 oscillation maximum while neutrinos from the other will arrive at the 1-3 oscillation minimum, greatly reducing the amplitude of the 1-3 oscillations whose observation is necessary to determine the hierarchy \cite{noi}.  Note that the neutrinos arriving from different reactors are not coherent, the interference effect results from the addition of probabilities and not wavefunctions.

Such an interference effect is present using just the reactors at Daya Bay and Ling Ao alone, because Daya Bay and Ling Ao I are separated by 1.1 km and Ling Ao II is 400 meters further.   Fortunately these reactors all lie more or less along a line, so a medium baseline detector perpendicular to this line will be the same distance from each reactor, eliminating the interference effect.  In Table \ref{inttab} we have determined the effect of this interference on the probability of success for a detector as a function of its distance from the center of mass of Daya Bay and Ling Ao, its angle with respect to this line and the number kton years of observations.  To illustrate the effect of the angle, the interference from other, more distant, reactors is not considered.  These will be included in the full simulations reported in Sec. \ref{rissez}.

\begin{table}[position specifier]
\centering
\begin{tabular}{c|l|l|l|l|l|l|l|l}
Baseline&kt yrs&$0^\circ$&$15^\circ$&$30^\circ$&$45^\circ$&$60^\circ$&$75^\circ$&$90^\circ$ \\
\hline\hline
50 km&60&\%&\%&\%&\%&\%&\%&\%\\
\hline
50 km&120&\%&\%&\%&\%&\%&\%&\%\\
\hline
58 km&60&\%&\%&\%&\%&\%&\%&\%\\
\hline
58 km&120&\%&\%&\%&\%&\%&\%&\%\\
\hline
\end{tabular}
\caption{The probability of determining the hierarchy as a function of baseline and number of kton years determined using the neural network.  The baseline is the distance from the center of mass of the Daya Bay, Ling Ao I and II reactor complexes considering the distances between these complexes.  The angles are measured with respect to the line that nearly passes through all three complexes.}
\label{inttab}
\end{table}

\section{Comparing Detector Locations Near Daya Bay} \label{rissez}

\subsection{Locations of Reactors and Detectors}

China's Guangdong province is among the best locations for a medium baseline reactor experiment because it contains the powerful reactor complex consisting of 2 reactors at Daya Bay and 4 at Ling Ao and yet it is free from the large reactor neutrino backgrounds caused by many smaller complexes in France and, modulo tsunami induced shutdowns, in Japan.  In the next few years a number of new reactor complexes will come on line in Guangdong.  Two distant complexes, TaiShan and YangJiang, will see their first reactors become operational already next year.  Two closer complexes, HuiDong and LuFeng, have already passed several critical steps in the approval process and may be built.  The relevant reactors are listed in Table \ref{reactortab}.

\begin{table}[position specifier]
\centering
\begin{tabular}{l|l|l|l|l}
Name&Status&Latitude&Longitude&Thermal Power\\
\hline\hline
Daya Bay&Operational&$22^\circ$ 35'  53" N&$114^\circ$ 32' 35" E& 5.8 GW\\
\hline
Ling Ao I&Operational&$22^\circ$ 36' 19" N& $114^\circ$  33' 4" E& 5.8 GW\\
\hline
Ling Ao II&Operational&$22^\circ$ 36' 31" N& $114^\circ$  33' 14" E& 5.8 GW\\
\hline
TaiShan I&Under Construction&$21^\circ$ 55' 9" N& $112^\circ$  58' 57" E& 9.2 GW\\
\hline
TaiShan II&Planned&$21^\circ$ 55'  N& $112^\circ$  59' E& 9.2 GW\\
\hline
YangJiang I&Under Construction&$21^\circ$ 42' 29" N& $112^\circ$  15' 32" E& 5.8 GW\\
\hline
YangJiang II&Under Construction&$21^\circ$ 42' 36" N& $112^\circ$  15' 41" E& 5.8 GW\\
\hline
YangJiang III&Planned&$21^\circ$ 43'  N& $112^\circ$  16'  E& 5.8 GW\\
\hline
HuiDong&Planned&$22^\circ$ 42'  N& $115^\circ$  0'  E& 17.4 GW\\
\hline
LuFeng&Planned&$22^\circ$ 45'  N& $115^\circ$  49'  E& 17.4 GW\\
\hline
\end{tabular}
\caption{Reactors in Guangdong}
\label{reactortab}
\end{table}

Depending on a detector location the new reactors may help or hinder the determination of the neutrino hierarchy.  They will help at each detector which is equidistant from two reactors,  as the fluxes add without interference.  This is the case in the proposed location of Ref \cite{yifangseminario} and also at our proposed location DongKeng.  If the two reactors are nearly at the same distance, as in the proposal of Ref. \cite{weihai} and our proposed location XiKengDing, the addition of the second reactor will increase the flux but interference between the detectors will decrease the amplitudes of the 1-3 oscillations.   

However in general the additional reactors will be so distant that the 1-3 oscillation peaks will be too close together to be distinguishable by a detector with resolution $\sigma_E$.  As a result, the additional reactors will simply supply a background which impedes the measurement of the hierarchy.

A medium baseline detector experiment may also be used to measure $\theta_{12}$.  This can be done by comparing the flux at the $1-2$ oscillation minima and maxima.  In the new reactor is twice as far away as the desired reactor, as is the case for most of the positions that we will consider, then the distant reactor will be at its 1-2 minimum at the same energy range in which neutrinos from the near reactor arrive at their 1-2 maximum.  This means that at the 1-2 minimum, where sensitivity to the hierarchy and to $\theta_{12}$ is maximized, the neutrino flux will be dominated by noise from the distant reactor \cite{noi}.  In general this makes the determination of both the hierarchy and $\theta_{12}$ more difficult.  However the 1-2 oscillations of distant reactors do lie within the resolution of the detector and so these can be used to gain more information about $\theta_{12}$.  Furthermore, if multiple detectors are considered, then they can break the degeneracy between $\theta_{12}$ and flux from distant reactors, allowing a more precise determination of both.

\begin{table}[position specifier]
\centering
\begin{tabular}{l|l|l|l|l|l|l|l|l}
Name&Altitude&Latitude (N) &Longitude (E) &DB&TS&YJ&HD&LF\\
\hline\hline
BaiYunZhang&1000 m&$22^\circ$ 53'  52"&$114^\circ$ 15' 14"& 44.5&170.5&244.1&79.5&161.6\\
\hline
ShiYaTou&500 m&$22^\circ$ 52'  14"&$114^\circ$ 17' 28"& 39.7&171.6&245.7&75.5&157.6\\
\hline
ShuangFeiJi&700 m&$22^\circ$ 54'  19"&$114^\circ$ 10' 0"& 51.6&164.2&237.0&88.3&170.6\\
\hline
SanJiaBi&600 m&$22^\circ$ 54'  8"&$114^\circ$ 10' 41"& 50.6&164.9&237.8&87.3&171.6\\
\hline
XiangTouShan&800 m&$23^\circ$ 15'  24"&$114^\circ$ 21' 0"& 75.4&205.0&275.3&90.4&160.9\\
\hline
BaiMianShi&400 m&$23^\circ$ 6'  27"&$114^\circ$ 37' 2"&  56.3&214.1&288.0&59.4&129.7\\
\hline
DongKeng&200 m&$22^\circ$ 6'  4"&$112^\circ$ 31' 9"& 216.5&51.8&51.0&264.1&347.3 \\
\hline
HuangDeDing&500 m&$22^\circ$ 5'  23"&$112^\circ$ 29' 55"& 218.9 &53.3&48.8&266.5&349.7\\
\hline
\end{tabular}
\caption{Potential locations for Daya Bay II detectors and distances to reactors in km. DB~indicates the distance to the weighted center of Daya Bay and Ling Ao I and II.}
\label{rivelatoretab}
\end{table}

In Table \ref{rivelatoretab} we list the detector positions that we propose together with BaiMianShi, which was proposed in Ref. \cite{weihai}.  These points are chosen to be at a medium baseline, to minimize interference effects, to be underneath mountains to minimize backgrounds and when possible to be a similar distance from multiple reactors to enhance the useful neutrino flux.  No site that we have found simultaneously achieves all of these goals.

\subsection{Monte Carlo Results}

The simulation of planned reactors is impeded by the fact that we do not know the configurations of the reactors.  These configurations are important when the reactors are used as sources of flux because it determines the reduction in the resulting 1-3 peaks due to interference.  It is not important when the reactors provide a background source.  While in general these configurations cannot be predicted, one natural guess is that the reactors will follow the coast, which is in general a sufficiently strong assumption to determine the resulting interference effect.

\begin{table}[position specifier]
\centering
\begin{tabular}{l|l|l|l|l|l|l|l|l}
Name&60 HD LF&60 HD&60 LF&60&120 HD LF&300 HD LF\\
\hline\hline
BaiYunZhang&77.8\%&&&&&\\
\hline
HS+XTS&&&&&&\\
\hline
SanJiaBii&77.0\%&&&&&\\
\hline
ShuangFeiJi&\%&&&&&\\
\hline
BaiMianShi&54.5\%&&&&&\\
\hline
BMS Ideal&79.4\%&&&&&\\
\hline
DongKeng&&&&&& \\
\hline
XiKengDing&&&&&&\\
\hline
\end{tabular}
\caption{Potential locations for Daya Bay II detectors and chances of success for various numbers of kton years of flux and choices of active reactors.  In all columns the reactors at Daya Bay, Ling Ao I and II, TaiShan and YangJiang.  HS+XTS refers to two detectors, one under HuaShan and one under XiangTouShan, each of which is half as large as indicated in the corresponding column.  BMS ideal is a mountainless point 2 km from BaiMianShi which is equidistant from the center of mass of the Daya Bay/Ling Ao complex and HuiDong.}
\label{risultatitab}
\end{table}

\section{Conclusions}

\section* {Acknowledgement}

\noindent
JE is supported by the Chinese Academy of Sciences
Fellowship for Young International Scientists grant number
2010Y2JA01. EC and XZ are supported in part by the NSF of
China.  


\end {document}

10 years ago Petcov and Piai suggested that a 20-25 km baseline neutrino detector may be able to determine the neutrino mass hierarchy if the solar mass splitting $\m21$ is greater than about $10^{-4}\ \mathrm{eV}^2$ \cite{petcovidea}.  It turned out that $\m21$ is smaller than this threshold, meaning that the difference between the normal and inverted hierarchy cannot be seen at a baseline below about 40 km.  The low fluxes at these long baselines led various groups \cite{hawaii,caojun} to conclude that the individual peaks in the neutrino spectrum would be difficult to resolve, and so the only hope for a hierarchy determination would be to sum them via a Fourier transform.  Even in this case the necessary detectors were extremely large and the experiments slow.

This all changed with the recent determination of $\theta_{13}$ \cite{daya,reno} confirming hints last year \cite{doublechooz,globale} that $\spp2213$ is up to 10 times higher than had been considered by the authors of Refs. \cite{hawaii,caojun}.  The larger mixing angle increases the sizes of the oscillations used to determine the hierarchy by an order of magnitude.  This means both that a reactor experiment to determine the hierarchy is now practical and also that the analysis of the optimal baseline and experimental configuration must now be redone.  Indeed, such a reanalysis is urgent as such an experiment will be built soon \cite{caojunseminario,renonuturn,yifangseminario}.

We will perform this updated analysis in two papers, mirroring the structure of Ref. \cite{caojun}.  While the second paper will include a description of an the results of our simulations, indicating for example the optimal baseline, in this first paper we will analyze a medium baseline reactor experiment analytically.  We will derive old observations relating observables of the Fourier transformed spectrum to the neutrino mass hierarchy and we will also provide new methods of determining various combinations of the neutrino masses and the hierarchy.  The optimal method will be a combination of the old and the new to be determined by simulations.  We will also discuss problems which have so far escaped attention in the literature.  

In particular we will discuss the fact that, since reactors are typically separated by of order 1 km in a reactor array, the baselines of neutrinos from different reactors are different.  As a result the oscillations of low energy neutrinos, which perform half an oscillation traveling from one reactor to the next, will be invisible at the detector as the neutrinos from one reactor arrive at the oscillation minimum while the other is at its maximum.  We will show that, due to a new degeneracy between the hierarchy and an effective mass difference, a determination of the neutrino mass hierarchy is impossible in an experiment which cannot resolve these low energy peaks.  Thus the angle between the detector and the lines extending between reactors is an essential variable in the determination of the optimal detector location.  For example, for a linear array of reactors like RENO or Daya Bay plus Ling Ao, this effect can be eliminated if the detector is placed orthogonal to the array.

The greatest sensitivity to the hierarchy arises near the 1-2 oscillation minimum, at about 60 km.  However the flux from a distance reactor at the 1-2 oscillation maximum of 120 km will dominate over the flux of the near reactor in this region.  In fact, this is will be the case for a detector placed 60 km away from Daya Bay and Ling Ao in the orthogonal direction, as the proposed Haifeng reactor will be near the 1-2 maximum.  Also if a detector is placed equidistant from Daya Bay and Haifeng at the position suggested in Ref. \cite{caojunseminario} then the 1-2 minimum neutrinos from Daya Bay and Haifeng will correspond to the 1-2 maximum for neutrinos from the proposed reactor at Lufeng \cite{yifangseminario}.  This not only makes the determination of the hierarchy more difficult, but also is detrimental to the sensitivity to $\theta_{12}$.  The ideal solution to this problem is to use two detectors at different distances, say 40 and 70 km.  However a more economical solution is to keep the baselines short so that the flux from the desired reactor complex dominates over the fluxes from others, for example one can consider a baseline of 45-50 km. 

We will begin in Sec. \ref{teorsez} with a review of standard results on 3-flavor neutrino oscillations and the electron antineutrino survival probability.  We describe the interference between the 1-3 and 2-3 oscillations \cite{petcovidea} which leads to beats and we numerically find the combined peaks.  Then in Sec. \ref{masssez} we describe how the positions of various peaks can be used to obtain various combinations of the neutrino mass differences.  Each peak provides a different combination.  And we will see that the first 10 peaks do not allow a determination of the mass hierarchy, but the next 5 do, which is why short baselines are not sufficient for a determination of the hierarchy.  In Sec. \ref{vincsez} we discuss the consequences of the finite energy resolution and the finite neutrino flux.  Both are provide obstacles to locating the $n>10$ peaks, and so for determining the hierarchy.  In Sec. \ref{fouriersez} we discuss analyses of the Fourier transform of the neutrino spectrum.  These have the advantage that, when the nonlinearity of the detector response is well understood, they sum the peaks together, and so render the signal stronger. We rederive three old ways in which the hierarchy can be determined from this transformed spectrum and add two new methods to the list. 

Finally in Sec. \ref{intsez} we discuss the consequences of the fact that not all of the neutrinos detected traveled the same distance.  The distances may differ by of order a kilometer because the individual reactors in an array are not coincident, leading to an interference effect which greatly diminishes the amplitudes of the low energy, high $n$, peaks. We will see that this interference can be avoided if the detector is placed perpendicular to the array.   Also, while it has long been known \cite{hawaii} that neutrinos from reactors at the 1-2 oscillation minimum baseline are the most useful for determining the hierarchy, we will see that this signal can be overwhelmed by neutrinos from distant reactors at the 1-2 maximum, and we see that this is indeed the case if a detector is placed at the 1-2 minimum orthogonal to the Daya Bay, Ling Ao reactor array.  This problem, as well as the related error in a determination of $\theta_{12}$, can be reduced by shortening the baseline or, if possible, adding another detector at a different baseline.

\section{The electron neutrino survival probability} \label{teorsez}

\subsection{Short and long oscillations}

The electron neutrino weak interaction eigenstate $|\nu_e\rangle$ is not an energy eigenstate $|k\rangle$, but it can be decomposed into a real sum of energy eigenstates
\beq
|\nu_e\rangle=\c12\c23|1\rangle+\s12\c13|2\rangle+\s13|3\rangle.
\eeq
In the relativistic limit, after traveling a distance $L$, the survival probability of a coherent electron (anti)neutrino wavepacket with energy $E$ can be expressed in terms of the mass matrix $\mathbf{M}$
\bea
P_{ee}&=&|\langle\nu_e|\mathrm{exp}\left(i\frac{\mathbf{M}^2L}{2E}\right)|\nu_e\rangle|^2\label{pee}\\
&=&\sp413+\cp412\cp413+\sp412\cp413+\frac{1}{2}(P_{12}+P_{13}+P_ {23})\nonumber\\
P_{12}&=&\spp2212\cp413\cos\left(\frac{\m21L}{2E}\right)\hsp
P_{13}=\cp212\spp2213\cos\left(\frac{\mn31L}{2E}\right)\nonumber\\
P_{23}&=&\sp212\spp2213\cos\left(\frac{\mn32L}{2E}\right)\nonumber
\eea
where $\m{i}{j}$ is the mass squared difference of mass eigenstates $i$ and $j$.  Notice that the survival probability is a sum of cosines and so its cosine Fourier transform with respect to the variable $L/E$ is just a sum of delta functions whereas its sine transform vanishes.

The three cosines in the survival probability (\ref{pee}) identify two characteristic frequencies of the $L/E$ oscillations.  The $P_{12}$ term oscillates at a low frequency 
\beq
\frac{\m21}{2}\sim 3.8\times 10^{-5}\ \mathrm{eV}^2\sim 0.17\ \mathrm{MeV/km} .
\eeq
Therefore the maximum 1-2 oscillation occurs at
\beq
\frac{L}{E}=\frac{\pi}{\m21/2}\sim 18 \ \mathrm{km/MeV}. \label{unodue}
\eeq
A medium baseline reactor experiment, with a baseline of under 100 km,  can observe at most one or two such oscillations.  Instead such experiments will focus on shorter oscillations, characterized by the $P_{13}$ term, which have frequency
\beq
\frac{\mn31}{2}\sim 1.2\times 10^{-3}\ \mathrm{eV}^2\sim 5.5\ \mathrm{MeV/km} 
\eeq
corresponding to a wavelength of
\beq
\Delta\left(\frac{L}{E}\right)=\frac{2\pi}{\mn31/2}\sim 1.1 \ \mathrm{km/MeV}. \label{corto}
\eeq
At a medium baseline reactor one may hope to see 5 to 15 such oscillations.

What about the $P_{23}$ term  in (\ref{pee})?  The frequency is $\mn32/2$, which is about 3\% more or less than $\mn31/2$ depending on the hierarchy.  However the amplitude is less than that of $P_{13}$ by a factor of
\beq
a=\tp212\sim 0.5.
\eeq
As the frequencies of the two short oscillations are similar but $P_{23}$ has a smaller amplitude, the total short distance oscillation $P_{13}+P_{23}$ is a deformation of $P_{13}$ alone.  However the 2-3 oscillations serve to slightly displace the 1-3 peaks and shift the amplitudes with a pattern which repeats at the beat frequency $\m12/2$.

More precisely, while the $n$th maximum of $P_{13}$ is at $L/E=4\pi n/\mn31$, the $n$th peak of $P_{13}+P_{23}$ is at
\beq
\frac{L}{E}=\frac{4\pi}{\mn31}(n+\an) \label{pichi}
\eeq
for a vector $\an$ which is determined entirely by the neutrino mass matrix.  As the derivatives of the neutrino spectrum and the 1-2 oscillations are small, the peaks of the total survival probability $P_{ee}$ and even its product with no-oscillation neutrino spectrum, which is the observed neutrino spectrum, are roughly located at the values given in Eq. (\ref{pichi}).

From Eq. (\ref{pichi}) it is possible to see the main obstruction to determinations of the hierarchy from near reactors.  The position of the $n$th peak is only sensitive to the mass combination
\beq
\Delta M^2_{eff}=\frac{\mn31}{1\pm\an/n}.
\eeq
As we will see, at low $n$, corresponding to peaks visible at relatively short baselines, the values of $\an$ are roughly linear.  Thus in this regime the positions of all of the peaks are sensitive to the same effective mass and so are independent of the hierarchy so long as that effective mass applies. {\textbf{The hierarchy can only be determined from the nonlinearity of $\an$.}}

\subsection{Finding $\an$}

The vector $\an$ encodes the effect of the neutrino mass matrix on the location of the survival probability peaks.  Therefore a knowledge of this function together with a measurement of the peaks allows one to reconstruct some elements of the mass matrix.

The values of $\an$ are determined from (\ref{pichi}) by the extrema of $P_{13}+P_{23}$
\bea
0&=&\frac{\partial}{\partial E}(P_{13}+P_{23})\propto \frac{\partial}{\partial E}\left[\cos\left(\frac{\mn31L}{2E}\right)+\tp212\cos\left(\frac{\mn32L}{2E}\right)\right]\\
&\propto&\sin\left(2\pi\an\right)+\left(1\pm\epsilon\right)\tp212\sin\left(2\pi[\an
\pm\epsilon(n+\an)]\right)\hsp
\epsilon=\frac{\m21}{\mn31}\nonumber
\eea
where the $-$ sign applies to the normal hierarchy and the $+$ sign to the inverse hierarchy.

As $\epsilon<<1$ and $\an<<n$ we may approximate
\beq
0\sim \sin(2\pi\an)+\tp212\sin(2\pi[\an\pm\epsilon n]). \label{alfeq}
\eeq
This can be expanded in a power series in $n$.  The highest energy peaks occur at small values of $n$, where the linear term in this expansion suffices
\beq
0\sim 2\pi(1+\tp212)\an\pm 2\pi\tp212\epsilon n
\eeq
which is easily solved for $\an$
\beq
\an\sim\mp\epsilon\sp212 n\sim \mp 0.011 n.
\eeq

Thus the n$th$ peak, for $n$ sufficiently small, lies at
\beq
\frac{L}{E}\sim\frac{4\pi}{\mn31}(1\mp\epsilon\sp212) n=\frac{4\pi n}{\mn31\pm\sp212\m21} \label{degen}
\eeq
where the lower sign corresponds to the normal hierarchy.  

The basic problem facing shorter baseline experiments, which are only sensitive to peaks at small $n$, is already apparent in Eq. (\ref{degen}).  The mass difference $\mn31$ is degenerate with the hierarchy
\beq
\mn31(\mathrm{direct})= \mn31(\mathrm{inverse})+2\sp212\m21 .
\eeq
Therefore any experiment with such a short baseline that all observable peaks have $n$ in the regime in which $\alpha$ is linear are, alone, incapable of determining the hierarchy.  They can, however, determine the combination 
\beq
\Delta M^2_{eff}=\mn31\pm\sp212\m21=\cp212\mn31+\sp212\mn32. \label{meff}
\eeq

Therefore reactor experiments can determine the hierarchy in only two ways.  Either an accurate determination of $\Delta M^2_{eff}$ can be combined with an accurate determination of another combination of the mass differences obtained from another experiment, a difference which needs to be known more precisely than the atmospheric mass difference measured by MINOS, or else the experiment needs to be sensitive to peaks at large enough $n$ that the linear approximation breaks down.  So just how large does $n$ need to be?

\subsection{Cubic terms is $\an$} \label{cubicosez}
To see where interference between 2-3 and 1-3 oscillations push the peaks of $P_{13}+P_{23}$ at larger values of $n$, we need to expand \an to cubic order
\beq
\an\sim  \mp\epsilon\sp212 n+b n^3. \label{cubico}
\eeq
and to substitute this expansion into Eq. (\ref{alfeq}).   The linear term in $\an$ already solves this equation at linear order.  At cubic order it yields
\beq
0\sim 2\pi(1+\tp212)b \pm\frac{4\pi^3}{3}\epsilon^3\sp212(\sp212-\cp212)       
\eeq
and so
\beq
b\sim\mp\frac{2\pi^2}{3}\epsilon^3\sp212\cp212(\sp212-\cp212) \sim \pm 2\times 10^{-5} .
\eeq

The linear approximation to $\an$ is reliable when the cubic term in Eq. (\ref{cubico}) is much smaller than the linear term
\beq
n<<\sqrt{\frac{\epsilon\sp212}{b}}\sim 20.
\eeq
For example, at the 10th peak the contribution of the cubic term to the energy is only one quarter of that of the linear term.  The linear term we have seen shifts the effective mass by about 1\%, thus the cubic term, which is the leading hierarchy-dependent term, only shifts the energy of the tenth peak by about one quarter of a percent.  The other hierarchy would lead to a shift of a quarter percent in the other direction, so overall the difference in the energies of the 10th peaks in the normal and inverted hierarchy is about one half percent.

Therefore if only the first ten peaks can be measured at a given baseline, the detector will need to be able to determine the position of the tenth peak with a precision of a half percent for only a one sigma determination of the hierarchy, making a determination of the hierarchy using such an experiment alone quite unlikely.

\begin{figure} 
\begin{center}
\includegraphics[width=5in,height=2in]{alfa.pdf}
\caption{$\an$ for the normal hierarchy}
\label{anfig}
\end{center}
\end{figure}

The cubic expansion is no longer reliable for higher peaks, but $\an$ can be determined numerically.  As seen in Fig. \ref{anfig} in the case of the normal hierarchy, it is periodic moduli $1/\epsilon$ and is zero at every multiple of $1/2\epsilon$.  But the main problem is that it is nearly linear for $n<10$, which is why the hierarchy is so nearly degenerate with the shift in the mass differences at all of these peaks.

A $2/k$ percent precision measurement of the peak energy of the the 14th peak would also give a $k$ sigma indication of the hierarchy.  The 14th peak can only be seen if it occurs at a high enough energy that the neutrino flux and detector resolution are sufficient to discern it.  For example if one requires it to appear at at least 3 MeV, so that the detector resolution may be better than 2\%, then using Eq. (\ref{corto}) the minimum baseline is about
\beq
L_{\mathrm{min}}\sim 45\ \mathrm{km}.
\eeq

\section{Determining mass differences from peak positions} \label{masssez}

\subsection{The reactor neutrino energy spectrum}

In the previous section we saw that the locations of the aperiodicity of the peaks is determined by the neutrino mass spectrum.  In particular, in $L/E$ space the peak positions are periodic modulo $1/\epsilon\sim 32$ and each set of $1/2\epsilon$ peaks is displaced about 1 percent either left or right depending on the hierarchy.  Thus, to determine the hierarchy, it suffices to measure the positions of enough peaks to within 2 percent precision.

The trouble with such a procedure is that no single experiment has access to all of the peaks, as one effectively has access to only a single baseline per detector and the energy spectrum is limited to that which is produced by a nuclear reactor, which effectively leads to a maximum usable neutrino energy.  Not even all of these energies are accessible as each type of neutrino detector has a minimum energy which it is able to detect.   To determine which peaks may be seen by a particular experiment, one must combine both of these constraints.

The neutrino flux from a reactor results almost entirely from decays of just 4 isotopes: \u35,\ \pu39,\ \pu41\ and \u38.  The flux $\phi_i(E)$ of neutrinos from each isotope $i$ is traditionally approximated as the exponential of a polynomial in the neutrino energy $E$ \cite{vogelengel}
\beq
\phi_i(E)=\mathrm{exp}\left(\sum_{k=1}^{m} a_{ki}E^{k-1}\right) \label{polinom}
\eeq
Theoretical errors on these fluxes are often claimed to be near the 2-3\% level, although recent theoretical fluxes \cite{nuovoflusso} appear to be about 6\% higher than the fluxes measured at very short baseline experiments \cite{reattoreanom} and at 1 kilometer experiments \cite{noiunokm}.  

The precision of such a phenomenological law depend on the degree $n$ of the fit polynomial.  For neutrinos with between 2 and 7.5 MeV, which will be the ones of interest in this note, it was shown in Ref. \cite{huber2004} that a quadratic ($m=3$) fit tends to introduce errors of order 2-3\% whereas a 6 parameter ($m=6$) fit introduces errors of order 1\%, well below the error in the theoretical flux.  The differences between these parameterizations are oscillations over a characteristic scale of 1-2 MeV, which are likely too broad to give a false signal for a 1-3 oscillation peak, but may well disguise the depth of 1-2 oscillations and so affect the measured value of $\theta_{12}$.  

The most recent theoretical estimate of the flux is Fig 53 of Ref. \cite{cartabianca}, which shows a systematic 3\% excess over last year's estimates \cite{nuovoflusso} at energies about 6 MeV and a one percent deficit beyond 4 MeV, which is well within the theoretical errors of the calculations.  This correction again will affect a single detector determination of $\theta_{12}$.

Not all of the neutrinos which are generated by the reactor will be measured.   The maximum number of neutrinos which can be measured at a given energy $E$ is the product of the produced flux with the fraction of neutrinos at that energy which can be measured.  For example, if neutrinos are measured via the inverse $\beta$ decay reaction
\beq
\overline{\nu}_e+p\rightarrow n+e^+
\eeq
then the maximum number of neutrinos detected is the flux/area at the baseline $L$ multiplied by the inverse $\beta$ decay cross section which at tree level is \cite{sezionedurto}
\beq
\sigma(E)=0.0952\times 10^{-42}\mathrm{cm}^2(E_e p_e/\mathrm{MeV}^2) \label{albero}
\eeq
where the positron energy and momenta are, ignoring the neutron recoil, given in terms of the neutron, proton and electron rest masses
\beq
E_e=E-m_n+m_p+m_e\sim E-780\ \mathrm{keV}\hsp
p_e=\sqrt{E_e^2-m_e^2}. \label{posenergia}
\eeq

\begin{figure} 
\begin{center}
\includegraphics[width=5in,height=2in]{Spettro_NO.pdf}
\caption{The theoretical reactor neutrino spectrum as measured with inverse beta decay.}
\label{flussoorig}
\end{center}
\end{figure}

Combining the reactor flux (\ref{polinom}), with the coefficients of Ref. \cite{vogelengel}, with the tree level cross section (\ref{albero}), one obtains the theoretical reactor flux, depicted in Fig. \ref{flussoorig}.  While inverse $\beta$ decay is kinematically forbidden if $E<m_n+m_e-m_p\sim 1.8$\ MeV, it can be seen in Fig. \ref{flussoorig} that the flux/energy is maximized at 3.6 MeV, falls to one third of its maximum by 6 MeV and to 10\% of its maximum by 7.5 MeV.  Thus useful information can only be obtained about the spectrum for energies within a factor of 2, which for each detector corresponds to values of $L/E$ within about a factor of 2.

\subsection{Effective masses at various baselines}

When neutrino oscillations are included, the theoretical flux $\Phi(E)$ from a reactor at a distance $L$ is then multiplied by the survival probability $P_{ee}$ given in Eq. (\ref{pee})
\beq
\Phi(E)=\sum_i c_i \phi_i(E)\sigma(E) P_{ee}(L/E)
\eeq
where $c_i$ is the quantity of each isotope in the reactor.

Using $\m21$ and $\spp2212$ from Ref. \cite{gando}, $\mn32$ determined by combining neutrino and antineutrino mass differences from Ref.\cite{minosneut2012}  and $\spp2213$ from \cite{noiunokm} 
\beq
\m21=7.50\times 10^{-5}\hspp
\mn32=2.41\times 10^{-3}\hspp
\spp2212=0.857\hspp
\spp2213=0.096
\eeq
where $\mn31$ is determined using the normal and inverted hierarchies, this total neutrino flux is given in Fig. \ref{tuttiflussi} at baselines of 40, 50, 60 and 70 km.

\begin{figure} 
\begin{center}
\includegraphics[width=3.2in,height=2in]{Plot_40_no_shift.pdf}
\includegraphics[width=3.2in,height=2in]{Plot_50_no_shift.pdf}
\includegraphics[width=3.2in,height=2in]{Plot_60_no_shift.pdf}
\includegraphics[width=3.2in,height=2in]{Plot_70_no_shift.pdf}
\caption{Theoretical neutrino fluxes, including 3 flavor oscillation, for both hierarchies as seen at 40, 50, 60 and 70 km.}
\label{tuttiflussi}
\end{center}
\end{figure}

The numbers $n$ of the local maxima can be read from (\ref{pichi}) by approximating $\mn31\sim\mn32$ and setting $\an=0$
\beq
n\sim\frac{4\pi}{\mn32}\frac{L}{E}\sim 0.9 \frac{L/\textrm{km}}{E/\textrm{MeV}}. \label{nvalore}
\eeq
As the error on $\mn32$ is of order 3\% and the difference between $\Delta M^2_{eff}$ and $\mn32$ is also of order 2\%, one may expect an error of 3-5\% in Eq. (\ref{nvalore}).  The fractional energy difference between the $n$th and $(n+1)$st peak is $1/n$.  Therefore an optimal detector can determine $n$ given a single peak only if $n$ is less than about 20, whereas an ordinary detector, ideally calibrated by radioactive decays close to the energy in question, can reliably determine the value $n$ of a peak when $n\leq 10$. 

Although the values of peaks in Fig \ref{tuttiflussi} are strongly dependent upon the hierarchy at every baseline, this does not mean that a measurement of the spectra can actually allow one to determine the hierarchy.  The problem, as was described in Sec. \ref{teorsez}, is that the energies of the first 10 or so peaks only determine the mass difference $\Delta M^2_{eff}$ of Eq (\ref{meff}).  

\begin{figure} 
\begin{center}
\includegraphics[width=3.2in,height=2in]{Plot_40_sovrapposto.pdf}
\includegraphics[width=3.2in,height=2in]{Plot_58_sovrapposto.pdf}
\caption{Theoretical neutrino fluxes, including 3 flavor oscillation, at 40 and 58 km for both hierarchies with the same value of $\Delta M^2_{eff}$.  In the left panel, at 40 km, the hierarchies are difficult to distinguish because $\an$ is nearly linear at the visible peaks.  This degeneracy is broken by the higher $n$ peaks visible at 58 km, seen in the right panel.}
\label{degenfig}
\end{center}
\end{figure}

Thus if the baseline is low enough so that only these peaks may be reliably measured, then there will be a value of $\Delta M^2_{eff}$ that reproduces the peaks for both hierarchies, as seen at 40 km in the first panel of Fig. \ref{degenfig}.   Here the peak $n=5$ can barely be seen at 6.6\ MeV, whereas $n=6$ at 5.5 MeV is clearly discernible.  The largest peaks are $n=7,\ 8,\ 9$ amd $10$.  However the energies of these peaks are independent of the hierarchy at constant $\Delta M^2_{eff}$.  The hierarchy-dependence becomes somewhat larger at the 11th peak, which is located at about 3.35 MeV in the case of the normal hierarchy and 3.30 MeV in the case of the inverted hierarchy.  Thus if $\Delta M^2_{eff}$  peaks can be determined at better than 1\% from the low $n$ peaks and then the location of the 11th peak can be determined with a precision of better than 1\%, a determination of the hierarchy would be barely possible.  The lower energy peaks are more smaller, but more hierarchy dependent.  For example, the 15th peak would be at 2.50 MeV with the normal hierarchy, but 2.42 MeV with the inverted hierarchy.  This 3\% difference is well within the resolution of the proposed detectors of Refs. (\cite{caojun}), and so when combined with an accurate measurement of $\Delta M^2_{eff}$ one may could potentially determine the hierarchy at the 1-2$\sigma$ level.



In the second panel one can see the electron neutrino survival probability at 58 km with both hierarchies and the same value of $\Delta M^2_{eff}$.  At this long baseline one can see maxima up to $n=20$, where the nonlinearity in $\an$ is appreciable and so, as one can see, the low and mid energy peaks are hierarchy-dependent at the 1-2\% level.  

\subsection{The 1-2 oscillation minimum}
It is clear from Fig. \ref{degenfig} that if $\Delta M^2_{eff}$ is determined from the low $n$ peaks then the locations of the peaks at the 1-2 oscillation minimum
\beq
n=\frac{1}{2\epsilon}\sim 15\hsp
E=\frac{L}{18\mathrm{\ km}}\textrm{MeV}
\eeq
are hierarchy-dependent, and so one may hope to use their locations to determine the hierarchy.  This can be done if the peak locations allow for a combination of the mass differences which is distinct from $\Delta M^2_{eff}$.  In fact, two such combinations can be determined, one from the location of the peaks and one from the distance between them.

To derive these two combinations, we will need to find $\an$  near the 1-2 oscillation minimum.  This is easily obtained by expanding (\ref{alfeq}) about $n=1/2\epsilon$
\beq
0\sim \sin(2\pi\an)-\tp212\sin\left(2\pi\left[\an\pm\epsilon \left(n-\frac{1}{2\epsilon}\right)\right]\right)
\eeq
where again the positive sign applies to the inverted hierarchy.  Linearly expanding the sine function yields
\beq
0\sim (1-\tp212)\an\mp\epsilon \left(n-\frac{1}{2\epsilon}\right)
\eeq
and so for $n\sim 1/2\epsilon$
\beq
\an\sim\pm\epsilon \frac{n-\frac{1}{2\epsilon}}{ 1-\tp212}.  \label{anminimo}
\eeq

At $n=1/2\epsilon$, $\an=0$.  Of course $n$ must be an integer, so it can never be precisely equal to $1/2\epsilon$.  Nonetheless, $\an$ will be over order $0.01$ when $n$ is at the closest integral value, leading to a less than 0.1\% contribution to the energy.  When $\an=0$, the energy of the peak is 
\beq
E=\frac{4\pi n L}{\mn31}=\frac{2\pi L}{\m21}
\eeq
allowing for a precise determination of $\m21$.  Of course, to know that one is at the 1-2 minimum by counting peaks, one must know $\epsilon$ and so this is related to a measurement of $\mn31$.  Alternately, one can find the 1-2 minimum by looking at the large oscillations in the flux due to 1-2 mixing and then count peaks to determine $n$ at the minimum, which allows for a determination of $\epsilon$ and so $\mn31$ from $\m21$.

The distance between the peaks near the 1-2 minimum allows for a determination of the mass differences which is independent of precise knowledge of the 1-2 oscillation parameters.  Inserting (\ref{anminimo}) in (\ref{pichi}) one finds that the distance between two peaks is
\beq
\Delta\left(\frac{L}{E}\right)=\frac{4\pi}{\mn31}\left(1\pm\frac{\epsilon}{1-\tp212}\right)
\eeq
and so it determines the effective mass
\beq
\Delta M^2_{min}=\left(1\mp\frac{\epsilon}{1-\tp212}\right)\mn31=\frac{2-\tp212}{1-\tp212}\mn31-\frac{1}{1-\tp212}\mn32 .
\eeq 

This effective mass is quite different from that defined in Eq. (\ref{meff}).  Approximating $\tp212=1/2$ one finds that the spacing between the first 10 peaks yields an effective mass
\beq
\Delta M^2_{eff}=\frac{2}{3}\mn31+\frac{1}{3}\mn32 \label{meffdue}
\eeq
while the peaks between $n=14$ and $n=18$ yield
\beq
\Delta M^2_{min}=3\mn31-2\mn32 . \label{mmindue}
\eeq
An detector at a baseline of less than 50 km can accurately determine the combination (\ref{meffdue}) while one with a baseline between 45 and 70 km, as it sees higher $n$ peaks, may more easily measure the combination (\ref{mmindue}).   The normal mass hierarchy is equivalent to the second mass being larger than the first, and so by comparing these masses one can determine the hierarchy.  

The difference between these masses is quite large, about 7\%, however a 7\% measurement of $\Delta M^2_{min}$ requires a 7\% precision in the measurement of the difference between the the peaks in the linear regime near to the 1-2 oscillation minimum $n=1/2\epsilon$.   As can be seen in Fig \ref{anfig}, this linear regime includes about 8 peaks, corresponding to a 40\% variation in energy.  Therefore a 7\% precision in the distance between the peaks requires a 4\% precision in the energy of the peaks, much less than was required at shorter baselines.  Therefore a comparison of the low $n$ peak positions and the 1-2 minimum peak separations is a promising test of the hierarchy, so long as the statistics are sufficient for the peaks to be observed.

\section{Resolution and flux constraints} \label{vincsez}

\subsection{Energy resolution}

The energy of the neutrino $E$ is determined from the positron energy $E_e$ by subtracting 780 keV (\ref{posenergia}).   The positron energy is determined by counting photoelectrons in a scintillator.  The number of photoelectrons is proportional to the positron energy, and so the energy resolution $\sigma_E$ is proportional to the square root of the positron energy.  For example, at Daya Bay II it has been suggested in Ref. \cite{caojun} that a resolution of
\beq
\sigma_E=0.03\sqrt{(E_e){\mathrm{MeV}}}.
\eeq
The observed positron  energy spectrum is then the convolution of the true spectrum with a Gaussian of width $\sigma$
\beq
P(E_e^{\textrm{observed}})=\int dE_e P(E_e) e^{-\frac{(E_e-E_e^{\textrm{observed}})^2}{2\sigma_E^2}}.
\eeq
Both spectra are displayed in Fig. \ref{smearfig}.

\begin{figure} 
\begin{center}
\includegraphics[width=3.2in,height=2in]{Smear_40.pdf}
\includegraphics[width=3.2in,height=2in]{Smear_50.pdf}
\includegraphics[width=3.2in,height=2in]{Smear_60.pdf}
\includegraphics[width=3.2in,height=2in]{Smear_70.pdf}
\caption{The true neutrino spectrum and the measured spectrum with a resolution of $3\%/E$ using the normal hierarchy at baselines of 40, 50, 60 and 70 km.  Notice that the lowest energy oscillations are smeared away and the energy threshold for this smearing increases with the baseline, such that at higher baselines the maximum $n$ observable increases slowly.}
\label{smearfig}
\end{center}
\end{figure}

Even the observed spectrum is never observed.  It is the best that can be hoped for, once the response of the detector is understood and with an infinite number of neutrinos.  Finite flux effects will be briefly discussed in Subsec. \ref{flussosez} and then discussed in detail in the companion paper on our simulation results.  A poorly understood detector response is fatal to the Fourier analysis that will be discussed in Sec. \ref{fouriersez}, but individual peaks may be analyzed where the response is understood.

Even in this idealized setting, it is clear from Fig. \ref{smearfig} that the low energy (high $n$) peaks cannot be resolved.  How high is $n$ for the biggest peak that can be resolved?  This depends on the baseline and the neutrino flux.  However a rough answer is obtained by asserting that the distance between the $n$th maximum and the adjacent minimum (\ref{corto}) 
\beq
\Delta E=0.9\frac{L/\mathrm{km}}{n}\mathrm{MeV}-0.9\frac{L/\mathrm{km}}{n+1/2}\mathrm{MeV}= 0.45 \frac{L/\mathrm{km}}{n^2}\mathrm{MeV}
\eeq
be greater than
\beq
2\sigma =0.06\sqrt{(E_e){\mathrm{MeV}}}=0.06\mathrm{\ MeV}\sqrt{\frac{L/\mathrm{km}}{n}-0.8}.
\eeq
These two quantities are equal when 
\beq
L/\mathrm{km}=8\times 10^{-3}n^3\left(1\pm\sqrt{1-\frac{225}{n^2}}\right). \label{lnecc}
\eeq
The equality with the positive sign yields the upper bound on observable $n$ for a given baseline $L$, whereas the other typically yields a value of $n$ at energies where no reactor neutrinos are observed.  

Alternately (\ref{lnecc}) provides the baseline $L$ necessary to observe the $n$th peak.  When $n\leq 15$ it provides no bound at all, so long as there is enough flux and the detector response is well understood, the energy resolution is not an obstruction for observing these peaks.  Of course at low baselines they may be difficult to observe because there simply are not many or any neutrinos observed at the corresponding energy.  At $n=16$, $17$ and $18$\ one finds minimum baselines of 45 km, 58 km and 72 km respectively.  Of course the true minimum depends on the neutrino flux.  But this rough estimate shows an essential point, that with a $3\%/\sqrt{E}$ fractional resolution any medium baseline, between about 40 and 70 km, is sufficient to observe the 1-2 oscillation minimum if there is enough neutrino flux.  Longer baselines only marginally extend the reach to higher peaks, although each of these peaks is in the 1-2 minimum region and so even just 1 or 2 more peaks can greatly enhance the possibility of determining the mass hierarchy.   

While maximum observable $n$ is reasonably independent of the baseline, this derivation shows that it is strongly dependent upon $\sigma_E$.  A resolution comparable to that of Daya Bay or RENO would lead to a maximum $n$ which is still in the linear region of $\an$, and so the hierarchy would be unobservable.

\subsection{How much neutrino flux is required?} \label{flussosez}

The neutrino flux that can be observed by a large, distant detector via inverse beta decay is still quite unknown, even the efficiency of the detector is difficult to predict.  A rough estimate can be made using the flux observed at Daya Bay, using the flux normalization-independent determination of $\t13$ to eliminate the loss due to 1-3 operation.  At Daya Bay 17.4 GW of thermal power yields 80 neutrinos/day at each 20 ton detector at a weighted baseline of 1600 meters.  Therefore the total measured flux/year from a $P$ GW reactor complex measured at a detector of mass $M$ at a baseline $L$ is roughly
\beq
\Phi=365\times 80\times \frac{P}{17.4\ \mathrm{GW}}\frac{M}{0.02\ \mathrm{ktons}}\frac{2.6\ \mathrm{km}^2}{L^2}=2.2\times 10^5\frac{(P/\mathrm{GW})(M/\mathrm{ktons})}{(L/\mathrm{km})^2}. \label{totflusso}
\eeq
For example, at a distance $L$ from the Daya Bay and Ling Ao reactors a 20 kton detector may observe $7.6\times 10^7/L^2$ neutrinos/year, where the length $L$ is measured in kilometers.

Let the energy width of a peak be $\Delta E$, the integrated flux within that energy range be $\phi$ and the peak be a fraction $A$ higher than the flux in that energy range from the reactor in question after 1-2 oscillations plus all other reactors.   In the absence of other reactors the fraction $f$ varies from $\spp2213$ at low $n$ to $4\spp2213$ at the 1-2 minimum.  The fractional error in the flux in that range will be $1/\sqrt{\phi}$.  Therefore the peak may be observed if $\sqrt{\phi} A>1$.  

Simply observing the peaks is useful for two reasons.  First of all, the distance between peaks is roughly $L\mn31/\pi$ and so by identifying peaks, one can check the consistency of the location of other peaks.  Second, recall that it is easier to determine the $n$ value of the well-separated, high energy, low $n$ peaks.  If one can observe the peaks from low to high $n$ then it is possible to count them and so determine the $n$ values of the low energy peaks.  While the hierarchy can be determined from the distance between the high $n$ peaks, as described in the previous subsection, without knowing the precise value of $n$, nonetheless if one knows the $n$ value of a high $n$ peak it can be used to determine a combination of the mass differences via (\ref{pichi}).  This determination has a precision of $1/n$, and so at high $n$ it leads to a precise determination.

However, to determine the hierarchy it is not enough to observe the peaks, one must determine their energies as precisely as possible.  How precisely may they be determined?  They may be determined within an energy $\delta E$ if the neutrino surplus can be seen in width $\delta E$ bands within the peak.  This requires
\beq
1<\sqrt{\phi}\frac{A\delta E}{\Delta E}=
\frac{\sqrt{\phi}Af}{n}
\eeq
where $f$ is the fractional energy precision desired.  This implies that the maximal fractional precision with which the energy of a peak can be measured by a detector with perfect resolution is
\beq
f>\frac{1}{An\sqrt{\phi}}.
\eeq
As mentioned above, with no unwanted backgrounds from other reactors, $A$ varies between $0.1$ at the 1-2 maxima to $0.4$ at the minima.  

Consider for example the tenth peak at a 40 km baseline, which lies at 3.6 MeV.  At this point $A\sim 0.2$.  The flux within the peak is about 5\% of the total flux (\ref{totflusso}), which each year at a 20 kton detector at 40 km from Daya Bay may be
\beq
\phi=0.05\times 7.6\times 10^7/(40)^2=2.4\times 10^3.
\eeq
Therefore, after $m$ years, the best resolution of an ideal detector would be
\beq
f_{min}=\frac{1}{0.2\times 10\sqrt{2.4m\times 10^3}}=\frac{1}{100\sqrt{m}}.
\eeq
Therefore an ideal detector can find the tenth peak energy to within $1/\sqrt{m}$ percent after 10 years.  As the peak width is greater than the resolution of Daya Bay or RENO, the high energy resolution proposed at a new medium baseline detector experiment in Ref. \cite{caojun} will not significantly alter the determination of this peak.  Therefore one may expect $\Delta M^2_{eff}$ to be determined to a precision significantly better than 1 percent at a 40 kilometer baseline experiment with no backgrounds from other reactors.

However, to determine the hierarchy, one also needs to determine another combination of the neutrino mass differences.  This requires the measurement of a higher $n$ peak.  Consider for example the $n=15$ peak at 2.5 MeV.  While it is in general a poor approximation to ignore the backgrounds provided by distant reactors, if one does ignore them then since since peak is near the 1-2 minimum, the relative peak height is $A=0.4$.   The total flux in the peak is only about 0.5
\beq
f_{min}=\frac{1}{0.4\times 15\sqrt{2.4m\times 10^2}}=\frac{1}{90\sqrt{m}}.
\eeq
As can be seen in Fig \ref{degenfig}, a 3\% precision required to determine the hierarchy.  As the width of the peak is about 3\%, a $3\%/\sqrt{E}$ resolution may reduce the amplitude by a factor of 2, still allowing for a determination of the hierarchy.

\section{The Fourier transform of the survival probability} \label{fouriersez}

While each peak provides some information regarding a combination of neutrino mass differences and therefore the hierarchy, it may well be that the fluxes are too weak or the backgrounds too small for the peaks to be reasonably well identified.  Complimentary information can be combined by combining the peaks.  As the electron survival probability is a sum of periodic cosine functions, they can be combined by a Fourier transform.  Even when individual peaks are hard to identify, the combination probed by the Fourier transform may well be visible and so may provide the best chance for determining the hierarchy \cite{hawaii}.

\subsection{The complex Fourier transform}

For simplicity we will approximate the observed electron antineutrino spectrum by a Gaussian distribution in $L/E$ space
\beq
\Phi\left(\frac{L}{E}\right)=e^{\left(\frac{L}{E}-L\langle\frac{1}{E}\rangle\right)^2/\sigma^2} \label{spec}
\eeq
where $\langle 1/E \rangle$ is the average $1/E$ of a neutrino which arrives.  While with only slightly more complicated equations the following could be avoided, we will make the crude approximation that (\ref{spec}) is the neutrino spectrum after 1-2 neutrino oscillations, and that the expectation value of the inverse energy is therefore taken with respect to the 1-2 oscillated spectrum, which depends upon $L$. 

1-3 oscillations affect this spectrum by introducing a modulation equal to $\Phi(L/E) P_{13}(L/E)$.  The Fourier transform of this modulation is
\bea
F_{13}(k)&=&\int d\left(\frac{L}{E}\right) \Phi(L/E)  P_{13}(L/E) e^{i\frac{kL}{E}}\\&=&\frac{\sigma\sqrt{\pi}\cp212}{4}\left(e^{\frac{\sigma^2}{4}\left(k+\frac{\mn31}{2}\right)^2}e^{i\left(k+\frac{\mn31}{2}\right)L\langle\frac{1}{E}\rangle}+(e^{\frac{\sigma^2}{4}\left(k.\frac{\mn31}{2}\right)^2}e^{i\left(k-\frac{\mn31}{2}\right)L\langle\frac{1}{E}\rangle}\right)\nonumber
\eea
where we have factored the $\spp2213$ out of the definition of $F_{13}$. This quantity has two peaks, one at $k=\mn31/2$ and one at $k=-\mn31/2$.  At each peak, one of the two terms in the parenthesis dominates, and the other is suppressed by a factor of order $e^{n^2/4}$, which is large enough that we will neglect the subdominant term.  Therefore, near the first peak
\beq
F_{13}(k)=\frac{\sigma\sqrt{\pi}\cp212}{4}e^{\frac{\sigma^2}{4}\left(k-\frac{\mn31}{2}\right)^2}e^{i\left(k-\frac{\mn31}{2}\right)L\langle\frac{1}{E}\rangle}
\eeq
The complex norm of $F$ has a maximum at $k=\mn31/2$, where $F$ is real.  The real part of $F$, corresponding to a cosine transform, is, within the validity of the approximations described above, symmetric about this maximum.  The imaginary part, corresponding to a sine transform, vanishes at this maximum and is antisymmetric about it.

The transform $F_{23}(k)$ can be calculated identically.  Near the positive $k$ maximum the sum is just
\bea
F(k)&=&\frac{\sigma\sqrt{\pi}}{4}\left[\cp212 e^{\frac{\sigma^2}{4}\left(k-\frac{\mn31}{2}\right)^2}e^{i\left(k-\frac{\mn31}{2}\right)L\langle\frac{1}{E}\rangle}+\sp212 e^{\frac{\sigma^2}{4}\left(k-\frac{\mn32}{2}\right)^2}e^{i\left(k-\frac{\mn32}{2}\right)L\langle\frac{1}{E}\rangle}\right]\nonumber\\
&=&\frac{\sigma\sqrt{\pi}}{4}\left[\cp212 e^{\frac{\sigma^2}{4}\left(k-\frac{\mn31}{2}\right)^2}+\sp212 e^{\frac{\sigma^2}{4}\left(k-\frac{\mn32}{2}\right)^2}e^{\pm i\frac{\m21 L}{2}\langle\frac{1}{E}\rangle}\right]e^{i\left(k-\frac{\mn31}{2}\right)L\langle\frac{1}{E}\rangle} \label{xform}
\eea
where the $+$ sign applies to the normal hierarchy.

The first term corresponds to the old peak, at $k=\mn31$ and the second to a new peak, which on its own would be $\tp212\sim 1/2$ as high as the first, as $k=\mn32$.  Of course, due to the phase difference of the two terms, for some choice of parameters the interference between the two terms implies that the second is not a local maximum of the norm of the Fourier transform.  However, each term is maximized when it is real, and so each term can be seen as a peak or a shoulder in the Fourier cosine transform.  The $k$ value of the peak then gives the corresponding mass difference.  Thus if the smaller peak is to the left, as smaller $k$, of the larger peak then $\mn32<\mn31$ and so one can conclude that there is a normal neutrino mass hierarchy \cite{hawaii}.

\subsection{Determining $\mn31$ at the 1-2 oscillation minimum}

We will refer to the baseline
\beq
L=\frac{2\pi}{\m21\langle\frac{1}{E}\rangle}
\eeq
as the 1-2 oscillation minimum, as it is roughly the baseline at which the largest fraction of neutrinos disappears as a result of $P_{12}$.   It is approximately 58 km.  At this distance the Fourier transform (\ref{xform}) simplifies as the two terms are precisely out of phase
\beq
F(k)=\frac{\sigma\sqrt{\pi}}{4}\left[\cp212 e^{\frac{\sigma^2}{4}\left(k-\frac{\mn31}{2}\right)^2}-\sp212 e^{\frac{\sigma^2}{4}\left(k-\frac{\mn32}{2}\right)^2}\right]e^{i\left(k-\frac{\mn31}{2}\right)\frac{2\pi}{\m21}} . \label{segno}
\eeq

The cosine transform is just the real part of $F$.  Up to $k$-independent factors, it is proportional to 
\beq
F_{\cos}(\tilde{k})=\left( e^{-\frac{\sigma^2}{4}\tilde{k}^2}-\tp212  e^{-\frac{\sigma^2}{4}\left(\tilde{k}\pm\frac{\m12}{2}\right)^2}\right)\cos\left(\frac{2\pi\tilde{k}}{\m21}\right) \label{coseq}
\eeq
where we have defined the distance to the $1-3$ peak to be
\beq
\tilde{k}=k-\frac{\mn31}{2}.
\eeq
The maxima of $F_{\cos}$ are found by setting to zero its derivative with respect to $\tilde{k}$, yielding the condition
\bea
&&\left(\frac{\sigma^2}{2}\tilde{k}+\frac{2\pi}{\m21}\tan\left(\frac{2\pi\tilde{k}}{\m21}\right)\right)e^{-\frac{\sigma^2}{4}\tilde{k}^2}\\&&\ \ \ \ \ \ \ \ \ \ \ \ \ =
\tp212\left(\frac{\sigma^2}{2}\left(\tilde{k}\pm\frac{\m21}{2}\right)+\frac{2\pi}{\m21}\tan\left(\frac{2\pi\tilde{k}}{\m21}\right)\right)e^{-\frac{\sigma^2}{4}\left(\tilde{k}\pm\frac{\m21}{2}\right)^2}\nonumber
\eea
where again the $+$ sign corresponds to the normal hierarchy and the $-$ sign to the inverted hierarchy.  The largest peak is the closest to $\tilde{k}=0$, the absolute maximum of the Fourier transform of $P_{13}$.  To find this peak we may expand $\tilde{k}$ about 0, at linear order we find
\beq
\tilde{k}=\pm\frac{\m21/2}{1+\frac{\pi^2}{2\beta}\left(e^\beta\ctp212-1\right)+2\beta} \label{ktilde}
\eeq
where we have defined the constant
\beq
\beta=\frac{\sigma^2\m21}{16}.
\eeq

As the denominator of (\ref{ktilde}) is much greater than 1, we learn that
\beq
|\tilde{k}|<<\frac{\m21}{2}
\eeq
and so the maximum of $F_{\cos}$ lies at approximately
\beq
k_{max}=\frac{\mn31}{2}.
\eeq
This means that if the detector is placed at the 1-2 oscillation minimum baseline then the peak of the cosine transform of the full neutrino spectrum lies essentially at the peak of $P_{13}$ alone.  This is not because because the frequencies of the $P_{13}$ and $P_{23}$ terms are similar, but because the absolute maximum of the Fourier transform of $P_{13}$, which corresponds to its frequency, happens to coincide with one of the minima of the cosine transform of $P_{23}$, which is not its frequency.  {\textbf{Thus, at the 1-2 oscillation minimum, the absolute maximum of the sum of two cosine transforms is coincident with that of $P_{13}$ alone, allowing for a direct and precise determination of $\mn31$}}.  This may be useful on its own even if the hierarchy has already been determined by NOvA of T2K.

\subsection{Determining the hierarchy at the 1-2 oscillation minimum}

Can this simplification also yield information about the hierarchy?  A precise determination of $\mn31$ combined with MINOS' best fit for $\mn32$ could give a 1$\sigma$ answer to this question, within 5 years MINOS+ may improve this to 2$\sigma$.  But with enough flux the peak structure Fourier transform alone yields some information about the hierarchy.  

\begin{figure} 
\begin{center}
\includegraphics[width=5.2in,height=2.5in]{FCT_58_D.pdf}
\caption{The cosine Fourier transform of $P_{12}$ (blue), $P_{23}$ (purple), $P_{13}$ (green) and $P_{ee}$ (yellow) at 58 km. {\textbf{This isn't quite true because our conventions for the P's are different, can we redo this figure with our conventions???}} Note that the $P_{13}$ and $P_{23}$ curves are just out of phase, so that the total extrema coincide with those of $P_{13}$.}
\label{cosfig}
\end{center}
\end{figure}

As can be seen in Fig. \ref{cosfig}, Just as the relative sign in (\ref{segno}) implied that the global maximum of the cosine transform of the total spectrum is essentially coincident with that of $P_{13}$, it also implies that the minima just to its left and right are coincident
\beq
k_{\min}^L=\frac{\mn31-\m21}{2}\hsp
k_{\min}^R=\frac{\mn31+\m21}{2}.
\eeq
These are minima for the simple reason that the cosine on the right of (\ref{coseq}) is equal to $-1$.  Substituting these values of $k$ into (\ref{coseq}) we find the values of the cosine transforms at the two minima
\beq
F_{\cos}(k_{min}^L)=-\left(e^{-\beta}-\tp212 e^{-\beta(-1\pm 1)^2}\right)\hsp
F_{\cos}(k_{min}^R)=-\left(e^{-\beta}-\tp212 e^{-\beta(1\pm 1)^2}\right).
\eeq
In the case of the normal hierarchy, the positive sign means that the second term is larger on the left, meaning that the minimum on the right is deeper.  In the case of the inverted hierarchy these two depths are interchanged, and so the peak on the left is deeper.  This is the criterion for determining the hierarchy from the cosine transform which was proposed in Ref. \cite{caojun}.

\begin{figure} 
\begin{center}
\includegraphics[width=5.2in,height=2.5in]{FST_58_D.pdf}
\caption{The sine Fourier transform of $P_{12}$ (blue), $P_{23}$ (purple), $P_{13}$ (green) and $P_{ee}$ (yellow) at 58 km. {\textbf{This isn't quite true because our conventions for the P's are different, can we redo this figure with our conventions???}} Note that the $P_{13}$ and $P_{23}$ curves are just out of phase, so that the total extrema coincide with those of $P_{13}$.}
\label{sinfig}
\end{center}
\end{figure}

A similar analysis can be applied to the imaginary part of $F(k)$, which is obtained via a sine transform.  As $F(\mn31/2)$ is real, the sine transform vanishes at the maximum of the global cosine transform.  The sine transform then has a maximum on the right and a minimum on the left.  The opposition of the phases of the Fourier transforms of $P_{13}$ and $P_{23}$ at the 1-2 oscillation minima again imply that these extrema of the sine transform of the full spectrum roughly coincide with the extrema of the sine transform of $P_{13}$ alone, as can be seen at 58 km in Fig. \ref{sinfig}.  They simply correspond to the values of $k$ for which the phase in (\ref{segno}) is $\pm\pi/2$
\beq
k^L=\frac{\mn31}{2}-\frac{\m21}{4}\hsp
k^R=\frac{\mn31}{2}+\frac{\m21}{4}.
\eeq
Defining the sine transform so as to take the imaginary part of $F$ with the same normalization as for the cosine transform one then finds
\beq
F_{\sin}(k^L)=-i\left(e^{-\beta/4}-\tp212 e^{-\frac{\beta}{4}(2\mp 1)^2}\right)\hsp
F_{\sin}(k^R)=i\left(e^{-\beta/4}-\tp212 e^{-\frac{\beta}{4}(2\pm 1)^2}\right)
\eeq
where the top sign corresponds to the normal hierarchy.  In the case of the normal hierarchy the second term in the minimum on the left is larger and so $|F_{\sin}(k^L)|<|F_{\sin}(k^R)|$, whereas in the case of the inverted hierarchy the maximum on the right is larger than the minimum on the left.  Thus we have reproduced the correlation between the hierarchy and the relative sizes of these extrema observed in Ref. \cite{caojun}.

Note that in the case of the normal hierarchy both extrema are more positive and in the case of the inverted hierarchy both are more negative, in other words {\textbf{near its maximum $F$ has a positive imaginary part in the case of a normal hierarchy and a negative imaginary part in the case of an inverted hierarchy}}.  This provides a new, third test which allows one to extrapolate the hierarchy from the Fourier transform of the survival probability.  Of course the optimal indicator of the hierarchy will be a weighted sum of these three tests, with weights that may be determined by series of simulations.  The ability to distinguish the hierarchies may also be improved by convoluting the observed spectrum with a slowly varying function of $(L/E)$ which weights neutrinos near the 1-2 minimum more heavily.  One can also use simulated data to determine which weighting functions are optimal for this task.

\subsection{Nonlinear Fourier transform}
The Fourier transform methods described above essentially work because the phase at the maximum is determined by the hierarchy, positive for the normal hierarchy and negative for the inverted hierarchy.  In the case of just $1-3$ oscillations, this phase is zero, since the corresponding oscillations are a cosine and the real part of the Fourier transform is determined by the cosine transform.  However when $P_{23}$ is included the peaks of the untransformed spectrum move according to Eq. (\ref{pichi}).  The distance between the untransformed peaks changes, which in general  moves the Fourier transformed peak.  

Critically, the untransformed $P_{13}+P_{23}$ also loses its mod $2/\mn31$ periodicity, as $\an$ is not a linear function of $n$.  A given detector is only sensitive to some of the peaks, corresponding to a certain region in Fig. \ref{anfig}.  Such regions are generally dominated by a domain in which $\an$ can be approximated not by a linear function, but by a linear function plus a constant offset.  This constant offset implies that the convolution of $P_{13}+P_{23}$ with the reactor neutrino spectrum is not of the form $\cos(\kappa L/E)$ but rather of the form $\cos(\kappa L/E+c)$ where the sign of $c$ is determined by the hierarchy.  This offset, $c$, leads to a translation in $L/E$ space which, after the Fourier transform, becomes a phase.  Thus the hierarchy determines an overall phase of the Fourier transform, which leads to the observable indicators described in the last section.

The Fourier transform method is robust since it sums multiple peaks together, and so it requires less neutrinos than a direct analysis of the positions of the peaks.  However it is inefficient because, as was just described, it works by approximating the function $\an$ to be affine in the energy range which is probed.  So one might ask if the performance would be increased by performing not an ordinary Fourier transform, but a Fourier transform with the nonlinearity of $\an$ built in.  The nonlinearity depends upon $\an$ which depends on the hierarchy and also weakly upon the neutrino mass matrix.  One may therefore attempt a nonlinear Fourier transform with both choices of hierarchy, and if desired, a weight function $g(L/E)$.  Such a nonlinear cosine transform is given by
\beq
F(k)=\int d\left(\frac{L}{E}\right)\Phi(L/E) P_{ee}(L/E) g(L/E) \cos\left(k\frac{L}{E}\pm 2\pi\alpha\left(\frac{k}{2\pi}\frac{L}{E}\right)\right)
\eeq
where $\alpha$ is a function which interpolates between the discrete values of $\an$ for the normal hierarchy and the positive sign corresponds to the inverted hierarchy.  This transform will add all of the peaks together with the same phase for the correct hierarchy whereas the peaks will be distorted by the other hierarchy.  Therefore the correct hierarchy can be determined from the fact that the corresponding nonlinear Fourier transform will have a larger absolute maximum.

\section{Interference effects} \label{intsez}

\subsection{Interference between reactors separated by 1 km} \label{unoproblema}

Consider two reactors of equal thermal power separated by a distance $D$.  If a detector is a distance $L>>D$ from the nearest and if the baseline makes an angle $\theta$ with respect to the line passing through both reactors, then the distance from the detector to the far reactor will be $L+d$ where $d=D\cos(\theta)$. 

Adding the flux from both reactors, one finds that the 1-3 oscillations interfere 
\beq
P_{13}\propto \cos\left(\frac{\mn31L}{2E}\right)+ \cos\left(\frac{\mn31(L+d)}{2E}\right)=2 \cos\left(\frac{\mn31d}{4E}\right)\cos\left(\frac{\mn31(L+d/2)}{2E}\right).
\eeq
The neutrinos from the two sources are not coherent, it is not the wavefunctions that add, but the probabilities.  And the result is that the amplitude of the oscillations at energy $E$ is suppressed by a factor of
\beq
\cos\left(\frac{\mn31d}{4E}\right)= \cos\left(3\frac{d/\mathrm{km}}{E/\mathrm{MeV}}\right).
\eeq
In particular they annihilate entirely when
\beq
\frac{d}{\mathrm{km}}=0.5 \frac{E}{\mathrm{MeV}}.
\eeq

Consider for example the pair of Daya Bay reactors and the two pairs of Ling Ao reactors which lie along a line at a distance of 0.8 and 1.3 km from the Daya Bay reactors.  The Daya Bay II reactor location suggested in Ref. \cite{caojunseminario} is at an angle of 15 degrees with respect to a continuation of this line, yielding $d=$1.2\ km for the nearest and furthest reactor pair.  As a result 1-3 oscillations at 2.4 MeV completely cancel between these two reactor pairs, leaving only the oscillations of the middle reactor, and so effectively damping the oscillation amplitude by a factor of 3, which implies that an equally precise measurement of a peak at that energy requires 9 times as much flux as it would have without interference.

At the peak energy, 3.6 MeV the annihilation is not complete, but is reduced by a factor of $\cos(1)=0.6$.  Thus the total amplitude of the peak from all three reactor pairs is reduced by about 30\%, and so twice the neutrino flux will be necessary to observe these peaks.

This problem can be avoided if the detector is equidistant from all of the reactors.  In the case of the Daya Bay and Ling Ao complex, in the case of RENO and somewhat trivially in the case of Double Chooz this is possible as all of the reactors essentially line along a line.  It means however that such a detector will not be equidistant from any reactor that may eventually be built at Haifeng, thus reducing the neutrino fluxes assumed in Ref. \cite{caojun} by a factor of 2.  Worse yet, the Haifeng reactor than provides a strong and undesirable background, as will be described in Subsec. \ref{centoproblema}.

\subsection{Interference between reactors separated by 100 km} \label{centoproblema}

The 1-2 oscillation minimum
\beq
L=\frac{2\pi E}{\mn31}=16\left(\frac{E}{\mathrm{MeV}}\right)\mathrm{km}
\eeq
provides an ideal baseline to determine the mass hierarchy for a number of reasons.  Among these is that while the $P_{13}+P_{23}$ oscillation amplitude of 
$(\cp212-\sp212)\spp2213$ is a factor of 3 less than its amplitude $\spp2213$ at the 1-2 maxima, the flux remaining after 1-2 oscillations is also smaller by a factor of $1/(1-\sp212\cp413)$ which is about 5.  Thus the 1-3 oscillations are a larger fraction of the total flux, making them easier to see above systematic errors, although not above statistical errors as $3>\sqrt{5}$.

However the smaller signal and smaller flux means that this part of the spectrum is particularly prone to interference from distant reactors, in particular those that may near the first 1-2 oscillation maximum
\beq
L=\frac{4\pi E}{\mn31}=32\left(\frac{E}{\mathrm{MeV}}\right)\mathrm{km}.
\eeq
In this case the addition factor of 5 in flux from the distant reactor outweighs the factor of 4 distance suppression, and so if both reactor complexes have the same strength than most of the flux at the 1-2 minimum is background.  This means that to obtain the same energy resolution at the 1-2 minimum peaks one will need more than 4 times as much neutrino flux.  

This is the case for example with a detector placed 58 km away from the Daya Bay/Ling Ao complex, perpendicular to the reactors.  It would be at the 1-2 maximum of the proposed Haifeng reactors and would also suffer significant contamination from the Taishan and Yanjiang reactor complexes, at each of which at least 3 reactors are already under construction.  Similarly a reactor placed 60 km from Daya Bay, Ling Ao and Haifeng would be at the 1-2 maximum of the proposed 17.4 GW thermal power reactor complex at Lufeng.

As this effect increases the 1-2 oscillation minimum flux, it is also a serious obstacle to an accurate measurement of $\theta_{12}$.  If a model of the background neutrino flux from distant reactors is wrong, the error in the total flux can be compensated for by an error in the determination of $\theta_{12}$.  Ideally this problem can be solved by using two medium baseline detectors instead of one, which would break this degeneracy.  However in a world limited by costs, it may instead be necessary to simply try to make the background from other reactors as small as possible, thus minimizing the potential error in the determination of the hierarchy and in the determination of $\theta_{12}$.

Unlike the short distance interference problem discussed in Subsec. \ref{unoproblema}, the fractional contamination caused by distance reactors depends on the baseline $L$ to the reactor whose neutrinos provide the signal.  The fractional contamination from undesired reactor neutrinos is inversely proportional to the desired signal strength, and so it is proportional to $L^2$.  Thus this problem is minimized by placing the detector as near to the reactor as possible.  If there is only one detector, and one wishes to use it to determine the hierarchy, then as explained in Subsec. \ref{cubicosez}, this minimum distance cannot be shorter than about 45 km.

\section{Conclusions}

The newly discovered high value of $\theta_{13}$ means that the determination of the neutrino mass hierarchy at a medium baseline reactor experiment is now practical.  Previous analysis of such experiments have assumed values of $\spp2213$ an order of magnitude or more below its true value.  At such low values, individual peaks of the electron antineutrino spectra could not be resolved and one instead needed to rely upon a Fourier analysis \cite{hawaii,caojun}.  

In this note we have reconsidered the determination of the neutrino mass hierarchy now that 1-3 oscillations are large and so individual 1-3 peaks in the spectrum can easily be observed.  We found that the position of each peak determines a particular combination of the neutrino mass differences, for example the first 10 peaks all determine the same difference $\cp212\mn31+\sp212\mn32$.  In particular this means that the first 10 peaks alone cannot be used to determine the hierarchy, as they are fit equally well by the wrong hierarchy model in which this mass difference is preserved.  On the other hand we found that the positions of the next 10 peaks depend on distinct combinations of the mass differences, and so combining the energies at different peaks with some beyond the $10$th can lead to a determination of the hierarchy.  We also estimated the detector resolution and neutrino flux which are needed for such a goal, although an accurate determination will be left for the simulations to be discussed in our companion paper.

The information about the hierarchy is therefore contained in the low energy peaks, which suffer the most from the effects of poor resolution,  low neutrino flux and interference.  While the individual peaks are somewhat difficult to resolve, the situation can nonetheless be improved with a Fourier transform.  We have analyzed the Fourier transform of the spectrum, deriving the phenomenological hierarchy indicators proposed in Refs. \cite{hawaii,caojun} and providing a new indicator, the complex phase at the peak of the Fourier transform, which we claim will be positive for the normal hierarchy and negative for the inverse hierarchy.  We also propose a new hierarchy-dependent nonlinear Fourier transform which will lead to a higher peak using the transform corresponding to the correct hierarchy.

The strength of the Fourier transform method is that it sums together all of the peaks to increase the strength of the signal with respect to the noise.  If the energy spectrum is shifted, this simply leads to an overall phase in the Fourier transform which does not seriously affect the analysis.  Likewise a uniform stretching of the spectrum simply shifts the peaks, which does not affect the determination of the hierarchy at all.  Any nonlinearity however poses a much more serious problem, as it can lead to an interference between the peaks in the spectrum which mutates or destroys the peak structure of the Fourier transform.  If the nature of the nonlinearity is known then one can adjust the analysis to correct for it \cite{petcov2010} however if the nonlinearity is known it could be corrected directly in the reading of the energy.  In general the nonlinearity of the response is only known at energies where the detector has been calibrated with radioactive sources.  It may therefore be desirable to modify the Fourier transform so as to weight the more reliable energies more heavily.  One may also wish to weight the energies near the 1-2 minimum more heavily, as it contributes few neutrinos but it is indispensable in a determination of the hierarchy.  These  weights can be optimized by testing various hierarchy determination algorithms against simulated data.

We also discussed two interference effects.  First of all, reactors in the same array are generally separated by distances of order 1 km, which means that neutrinos arriving at a detector from one reactor at a 1-3 maximum may be at the same energy as those arriving from another at a 1-3 minimum.  The result is that the 1-3 oscillation signal can be severely reduced at the low energies in which this oscillation can occur within a reactor complex.  These are just the low energies which are necessary for the determination of the hierarchy, and so this interference poses a serious problem.  One solution is to place the detectors orthogonal to arrays of reactors.

The second interference effect arises from the fact that while the 1-2 oscillation minimum is the most useful energy range at which to determine the hierarchy, it enjoys a much lower neutrino flux than the 1-2 oscillation maximum.  As a result at these energy ranges one can expect serious contamination from distant reactors at their 1-2 maximum.  This problem, as well as the degeneracy between the neutrino flux from distant reactors and $\theta_{12}$, is optimally solved by using two detectors at different baselines.  However a cheaper solution to the same problem is to use a single detector at a shorter baseline, such as 45-50 km.


\section* {Acknowledgement}

\noindent
JE is supported by the Chinese Academy of Sciences
Fellowship for Young International Scientists grant number
2010Y2JA01. EC and XZ are supported in part by the NSF of
China.  


\end{document}

\bibitem{lsnd}
A.~Aguilar-Arevalo {\it et al.}  [LSND Collaboration],
  ``Evidence for neutrino oscillations from the observation of anti-neutrino(electron) appearance in a anti-neutrino(muon) beam,''
  Phys.\ Rev.\ D {\bf 64} (2001) 112007
  [hep-ex/0104049].

\bibitem{minibooneanom}
A.~A.~Aguilar-Arevalo {\it et al.}  [The MiniBooNE Collaboration],
  ``A Search for electron neutrino appearance at the $\Delta m^{2} \sim 1$eV$^{2}$ scale,''
  Phys.\ Rev.\ Lett.\  {\bf 98} (2007) 231801
  [arXiv:0704.1500 [hep-ex]].
A.~A.~Aguilar-Arevalo {\it et al.}  [MiniBooNE Collaboration],
  ``Unexplained Excess of Electron-Like Events From a 1-GeV Neutrino Beam,''
  Phys.\ Rev.\ Lett.\  {\bf 102} (2009) 101802
  [arXiv:0812.2243 [hep-ex]].
 A.~A.~Aguilar-Arevalo {\it et al.}  [The MiniBooNE Collaboration],
  ``Event Excess in the MiniBooNE Search for $\bar \nu_\mu \rightarrow \bar \nu_e$ Oscillations,''
  Phys.\ Rev.\ Lett.\  {\bf 105} (2010) 181801
  [arXiv:1007.1150 [hep-ex]].

\bibitem{minosanom}
P.~Adamson {\it et al.}  [MINOS Collaboration],
  ``First direct observation of muon antineutrino disappearance,''
  Phys.\ Rev.\ Lett.\  {\bf 107} (2011) 021801
  [arXiv:1104.0344 [hep-ex]].

\bibitem{zichichi}
A.~Zichichi,
``Results from LVD-OPERA Combined Analysis: A Time-Shift in the OPERA Setup,"
available online at http://agenda.infn.it/getFile.py/access?resId=0\&materialId=slides\&confId=4896.

\bibitem{miniboonetuttobene}
E.~D.~Zimmerman [MiniBooNE Collaboration],
  ``Updated Search for Electron Antineutrino Appearance at MiniBooNE,''
  arXiv:1111.1375 [hep-ex].

\bibitem{minostuttobene}
P.~Adamson {\it et al.}  [MINOS Collaboration],
  ``An improved measurement of muon antineutrino disappearance in MINOS,''
  arXiv:1202.2772 [hep-ex].

\bibitem{nuovoflusso}
T.~.A.~Mueller, D.~Lhuillier, M.~Fallot, A.~Letourneau, S.~Cormon, M.~Fechner, L.~Giot and T.~Lasserre {\it et al.},
  ``Improved Predictions of Reactor Antineutrino Spectra,''
  Phys.\ Rev.\ C {\bf 83} (2011) 054615
  [arXiv:1101.2663 [hep-ex]].
P.~Huber,
  ``On the determination of anti-neutrino spectra from nuclear reactors,''
  Phys.\ Rev.\ C {\bf 84} (2011) 024617
   [Erratum-ibid.\ C {\bf 85} (2012) 029901]
  [arXiv:1106.0687 [hep-ph]].

\bibitem{reattoreanom}
G.~Mention, M.~Fechner, T.~.Lasserre, T.~.A.~Mueller, D.~Lhuillier, M.~Cribier and A.~Letourneau,
  ``The Reactor Antineutrino Anomaly,''
  Phys.\ Rev.\ D {\bf 83} (2011) 073006
  [arXiv:1101.2755 [hep-ex]].

\bibitem{smirnov}
P.~C.~de Holanda and A.~Y.~.Smirnov,
  ``Homestake result, sterile neutrinos and low-energy solar neutrino experiments,''
  Phys.\ Rev.\ D {\bf 69} (2004) 113002
  [hep-ph/0307266].
P.~C.~de Holanda and A.~Y.~.Smirnov,
  ``Solar neutrino spectrum, sterile neutrinos and additional radiation in the Universe,''
  Phys.\ Rev.\ D {\bf 83} (2011) 113011
  [arXiv:1012.5627 [hep-ph]].

\bibitem{icecube}
R.~Abbasi {\it et al.}  [IceCube Collaboration],
  ``Measurement of the atmospheric neutrino energy spectrum from 100 GeV to 400 TeV with IceCube,''
  Phys.\ Rev.\ D {\bf 83} (2011) 012001
  [arXiv:1010.3980 [astro-ph.HE]].

\bibitem{giuntireview}
  C.~Giunti and M.~Laveder,
  ``Implications of 3+1 Short-Baseline Neutrino Oscillations,''
  Phys.\ Lett.\ B {\bf 706} (2011) 200
  [arXiv:1111.1069 [hep-ph]].

\bibitem{sterilecosm}
J.~Hamann, S.~Hannestad, G.~G.~Raffelt and Y.~Y.~Y.~Wong,
  ``Sterile neutrinos with eV masses in cosmology: How disfavoured exactly?,''
  JCAP {\bf 1109} (2011) 034
  [arXiv:1108.4136 [astro-ph.CO]].

\bibitem{dayabay}
  F.~P.~An {\it et al.}  [DAYA-BAY Collaboration],
  ``Observation of electron-antineutrino disappearance at Daya Bay,''
  Phys.\ Rev.\ Lett.\  {\bf 108} (2012) 171803
  [arXiv:1203.1669 [hep-ex]].

\bibitem{piureattori}
L.~A.~Mikaelyan and V.~V.~Sinev,
  ``Neutrino oscillations at reactors: What next?,''
  Phys.\ Atom.\ Nucl.\  {\bf 63} (2000) 1002
   [Yad.\ Fiz.\  {\bf 63N6} (2000) 1077]
  [hep-ex/9908047].

\bibitem{neut2012}
D. Dwyer,
``Daya Bay Results," presented at Neutrino 2012 in Kyoto.
Soon to be available at http://neu2012.kek.jp/neu2012/programme.html.

\bibitem{doublechooz}
Y.~Abe {\it et al.}  [DOUBLE-Chooz Collaboration],
  ``Indication for the disappearance of reactor electron antineutrinos in the Double Chooz experiment,''
  Phys.\ Rev.\ Lett.\  {\bf 108} (2012) 131801
  [arXiv:1112.6353 [hep-ex]].

\bibitem{reno}
  J.~K.~Ahn {\it et al.}  [RENO Collaboration],
  ``Observation of Reactor Electron Antineutrino Disappearance in the RENO Experiment,''
  Phys.\ Rev.\ Lett.\  {\bf 108} (2012) 191802
  [arXiv:1204.0626 [hep-ex]].

\bibitem{nuturn}
``Observation of reactor neutrino disappearance at RENO," presented at $\nu$TURN 2012 under Gran Sasso.  Available at http://agenda.infn.it/contributionListDisplay.py?confId=4722.

\bibitem{globale1}
G.~L.~Fogli, E.~Lisi, A.~Marrone, A.~Palazzo and A.~M.~Rotunno,
  ``Evidence of $\theta_{13}$>0 from global neutrino data analysis,''
  Phys.\ Rev.\ D {\bf 84} (2011) 053007
  [arXiv:1106.6028 [hep-ph]].

\bibitem{globale2}
  T.~Schwetz, M.~Tortola and J.~W.~F.~Valle,
  ``Where we are on $\theta_{13}$: addendum to 'Global neutrino data and recent reactor fluxes: status of three-flavour oscillation parameters',''
  New J.\ Phys.\  {\bf 13} (2011) 109401
  [arXiv:1108.1376 [hep-ph]].

\bibitem{paloverde}
F.~Boehm, J.~Busenitz, B.~Cook, G.~Gratta, H.~Henrikson, J.~Kornis, D.~Lawrence and K.~B.~Lee {\it et al.},
  ``Final results from the Palo Verde neutrino oscillation experiment,''
  Phys.\ Rev.\ D {\bf 64} (2001) 112001
  [hep-ex/0107009].

\bibitem{chooz}
M.~Apollonio {\it et al.}  [Chooz Collaboration],
  ``Search for neutrino oscillations on a long baseline at the Chooz nuclear power station,''
  Eur.\ Phys.\ J.\ C {\bf 27} (2003) 331
  [hep-ex/0301017].

\bibitem{neutdarke}
X.~-J.~Bi, P.~-H.~Gu, X.~-l.~Wang and X.~-M.~Zhang,
  ``Thermal leptogenesis in a model with mass varying neutrinos,''
  Phys.\ Rev.\ D {\bf 69} (2004) 113007
  [hep-ph/0311022].
  R.~Takahashi and M.~Tanimoto,
  ``Model of mass varying neutrinos in SUSY,''
  Phys.\ Lett.\ B {\bf 633} (2006) 675
  [hep-ph/0507142].
  R.~Takahashi and M.~Tanimoto,
  ``Speed of sound in the mass varying neutrinos scenario,''
  JHEP {\bf 0605} (2006) 021
  [astro-ph/0601119].
  E.~Ciuffoli, J.~Evslin, J.~Liu and X.~Zhang,
  ``OPERA and a Neutrino Dark Energy Model,''
  arXiv:1109.6641 [hep-ph].

\bibitem{neal04}
  D.~B.~Kaplan, A.~E.~Nelson and N.~Weiner,
  ``Neutrino oscillations as a probe of dark energy,''
  Phys.\ Rev.\ Lett.\  {\bf 93} (2004) 091801
  [hep-ph/0401099].

\bibitem{tortola}
  M.~Tortola, J.~W.~F.~Valle and D.~Vanegas,
  ``Global status of neutrino oscillation parameters after recent reactor measurements,''
  arXiv:1205.4018 [hep-ph].

\bibitem{foglinuovo}
G.L. Fogli, E. Lisi, A. Marrone, D. Montanino, A. Palazzo and A.M. Rotunno,
``Global analysis of neutrino masses, mixings and phases: entering the era of leptonic CP violation searches,"
arXiv:1205.5254 [hep-ph].

\bibitem{bugey4}
Y.~Declais, H.~de Kerret, B.~Lefievre, M.~Obolensky, A.~Etenko, Y.~.Kozlov, I.~Machulin and V.~Martemyanov {\it et al.},
  ``Study of reactor anti-neutrino interaction with proton at Bugey nuclear power plant,''
  Phys.\ Lett.\ B {\bf 338} (1994) 383.

\bibitem{dayafeb}
F.~P.~An {\it et al.}  [Daya Bay Collaboration],
  ``A side-by-side comparison of Daya Bay antineutrino detectors,''
  arXiv:1202.6181 [physics.ins-det].

\bibitem{daya2007}
  X.~Guo {\it et al.}  [Daya-Bay Collaboration],
  ``A Precision measurement of the neutrino mixing angle theta(13) using reactor antineutrinos at Daya-Bay,''
  hep-ex/0701029.


\bibitem{tredueterm}
 A.~Melchiorri, O.~Mena, S.~Palomares-Ruiz, S.~Pascoli, A.~Slosar and M.~Sorel,
  ``Sterile Neutrinos in Light of Recent Cosmological and Oscillation Data: A Multi-Flavor Scheme Approach,''
  JCAP {\bf 0901} (2009) 036
  [arXiv:0810.5133 [hep-ph]].

\bibitem{neutrinoasym}
  S.~Hannestad, I.~Tamborra and T.~Tram,
  ``Thermalisation of light sterile neutrinos in the early universe,''
  arXiv:1204.5861 [astro-ph.CO].

\bibitem{baoscoperta}
  D.~J.~Eisenstein {\it et al.}  [SDSS Collaboration],
  ``Detection of the baryon acoustic peak in the large-scale correlation function of SDSS luminous red galaxies,''
  Astrophys.\ J.\  {\bf 633} (2005) 560
  [astro-ph/0501171].


\bibitem{bao}
W.~J.~Percival, S.~Cole, D.~J.~Eisenstein, R.~C.~Nichol, J.~A.~Peacock, A.~C.~Pope and A.~S.~Szalay,
  ``Measuring the Baryon Acoustic Oscillation scale using the SDSS and 2dFGRS,''
  Mon.\ Not.\ Roy.\ Astron.\ Soc.\  {\bf 381} (2007) 1053
  [arXiv:0705.3323 [astro-ph]].

\bibitem{cosmomc}
  A.~Lewis and S.~Bridle,
  ``Cosmological parameters from CMB and other data: A Monte Carlo approach,''
  Phys.\ Rev.\ D {\bf 66} (2002) 103511
  [astro-ph/0205436].

\bibitem{wmap}
E.~Komatsu {\it et al.}  [WMAP Collaboration],
  ``Seven-Year Wilkinson Microwave Anisotropy Probe (WMAP) Observations: Cosmological Interpretation,''
  Astrophys.\ J.\ Suppl.\  {\bf 192} (2011) 18
  [arXiv:1001.4538 [astro-ph.CO]].

\bibitem{Suzuki:2011hu}
  N.~Suzuki {\it et al.},
  ``The Hubble Space Telescope Cluster Supernova Survey: V. Improving the Dark
  Energy Constraints Above z>1 and Building an Early-Type-Hosted Supernova
  Sample,''
  arXiv:1105.3470 [astro-ph.CO].

\bibitem{h}
A.~G.~Riess, L.~Macri, S.~Casertano, H.~Lampeitl, H.~C.~Ferguson, A.~V.~Filippenko, S.~W.~Jha and W.~Li {\it et al.},
  ``A 3\% Solution: Determination of the Hubble Constant with the Hubble Space Telescope and Wide Field Camera 3,''
  Astrophys.\ J.\  {\bf 730} (2011) 119
   [Erratum-ibid.\  {\bf 732} (2011) 129]
  [arXiv:1103.2976 [astro-ph.CO]].

\bibitem{nessundivergenza}
J.~-Q.~Xia, G.~-B.~Zhao, B.~Feng, H.~Li and X.~Zhang,
  ``Observing dark energy dynamics with supernova, microwave background and galaxy clustering,''
  Phys.\ Rev.\ D {\bf 73} (2006) 063521
  [astro-ph/0511625].
G.~-B.~Zhao, J.~-Q.~Xia, M.~Li, B.~Feng and X.~Zhang,
  ``Perturbations of the quintom models of dark energy and the effects on observations,''
  Phys.\ Rev.\ D {\bf 72} (2005) 123515
  [astro-ph/0507482].

\bibitem{altroquintom}
  E.~Giusarma, M.~Archidiacono, R.~de Putter, A.~Melchiorri and O.~Mena,
  ``Sterile neutrino models and nonminimal cosmologies,''
  Phys.\ Rev.\ D {\bf 85} (2012) 083522
  [arXiv:1112.4661 [astro-ph.CO]].

\bibitem{unione2}
N.~Suzuki, D.~Rubin, C.~Lidman, G.~Aldering, R.~Amanullah, K.~Barbary, L.~F.~Barrientos and J.~Botyanszki {\it et al.},
  ``The Hubble Space Telescope Cluster Supernova Survey: V. Improving the Dark Energy Constraints Above z=1 and Building an Early-Type-Hosted Supernova Sample,''
  Astrophys.\ J.\  {\bf 746} (2012) 85
  [arXiv:1105.3470 [astro-ph.CO]].

\bibitem{quintom}
B.~Feng, M.~Li, Y.~-S.~Piao and X.~Zhang,
  ``Oscillating quintom and the recurrent universe,''
  Phys.\ Lett.\ B {\bf 634} (2006) 101
  [astro-ph/0407432].
  B.~Feng, X.~L.~Wang and X.~M.~Zhang, 
``Dark energy constraints from the cosmic age and supernova,''
 Phys.\ Lett.\  B {\bf 607} (2005) 35  
 [arXiv:astro-ph/0404224].
  X.~-F.~Zhang, H.~Li, Y.~-S.~Piao and X.~-M.~Zhang,
``Two-field models of dark energy with equation of state across -1,''
  Mod.\ Phys.\ Lett.\ A {\bf 21}, 231 (2006)
  [astro-ph/0501652].

\bibitem{neutrinoasymvecchio}
  R.~Foot and R.~R.~Volkas,
  ``Reconciling sterile neutrinos with big bang nucleosynthesis,''
  Phys.\ Rev.\ Lett.\  {\bf 75} (1995) 4350
  [hep-ph/9508275].

\bibitem{fr}
  H.~Motohashi, A.~A.~Starobinsky and J.~'i.~Yokoyama,
  ``Cosmology based on f(R) Gravity admits 1 eV Sterile Neutrinos,''
  arXiv:1203.6828 [astro-ph.CO].

\bibitem{robert}
R.~H.~Brandenberger, N.~Kaiser, D.~N.~Schramm and N.~Turok,
  ``Galaxy and Structure Formation with Hot Dark Matter and Cosmic Strings,''
  Phys.\ Rev.\ Lett.\  {\bf 59} (1987) 2371.
R.~H.~Brandenberger, A.~Mazumdar and M.~Yamaguchi,
  ``A Note on the robustness of the neutrino mass bounds from cosmology,''
  Phys.\ Rev.\ D {\bf 69} (2004) 081301
  [hep-ph/0401239].

\bibitem{monopoli}
J.~Evslin and S.~B.~Gudnason,
  ``High Q BPS Monopole Bags are Urchins,''
  arXiv:1111.3891 [hep-th].
J.~Evslin and S.~B.~Gudnason,
  ``Dwarf Galaxy Sized Monopoles as Dark Matter?,''
  arXiv:1202.0560 [astro-ph.CO].

\bibitem{lsndnonstandard}
  G.~Karagiorgi, M.~H.~Shaevitz and J.~M.~Conrad,
  ``Confronting the short-baseline oscillation anomalies with a single sterile neutrino and non-standard matter effects,''
  arXiv:1202.1024 [hep-ph].

\bibitem{envir}
  G.~Karagiorgi, M.~H.~Shaevitz and J.~M.~Conrad,
  ``Confronting the short-baseline oscillation anomalies with a single sterile neutrino and non-standard matter effects,''
  arXiv:1202.1024 [hep-ph].

\end{thebibliography}

\end{document}